\documentclass[prd, twocolumn, lengthcheck,letterpaper,nofootinbib,showpacs]{revtex4-1}

\usepackage{color}
\usepackage{graphicx, subfigure}
\usepackage{float}
\usepackage{amsmath}
\usepackage{acronym}
\usepackage{multirow}
\usepackage{amsfonts}
\usepackage{ulem}

\usepackage{color}

\newcommand{\beq}{\begin{equation}}
\newcommand{\eeq}{\end{equation}}
\newcommand{\bea}{\begin{eqnarray}} 
\newcommand{\eea}{\end{eqnarray}}
\newcommand{\ba}{\begin{array}}
\newcommand{\ea}{\end{array}}

\newcommand{\degree}{\ensuremath{^\circ}}

\newcommand{\Msun}{M_{\odot}}

\newcommand{\mytextrm}[1]{{}}

\newcommand{\quarter}{\frac{1}{4}}
\newcommand{\half}{\frac{1}{2}}

\newcommand{\Degs}{\ensuremath{^\circ}}

\def\ltsima{$\; \buildrel < \over \sim \;$}
\def\simlt{\lower.5ex\hbox{\ltsima}}
\def\gtsima{$\; \buildrel > \over \sim \;$}
\def\simgt{\lower.5ex\hbox{\gtsima}}
\begin{document}

\title{The effect of small inter-pulsar distance variations in stochastic gravitational wave background searches with Pulsar Timing Arrays }


\author{Chiara~M.~F.~Mingarelli}
\email{chiara@caltech.edu}
\affiliation{School of Physics and Astronomy, University of Birmingham, Edgbaston, Birmingham B15 2TT, UK}
\affiliation{Theoretical Astrophysics, California Institute of Technology, 1200 E California Blvd., M/C 350-17, Pasadena, CA 91125, USA }
\affiliation{Max Planck Institute for Radio Astronomy, 
Auf dem H\"{u}gel 69, D-53121 Bonn, Germany }

\author{Trevor Sidery}
\affiliation{School of Physics and Astronomy, University of Birmingham, Edgbaston, Birmingham B15 2TT, UK}
\date\today

\begin{abstract} 
One of the primary objectives for Pulsar Timing Arrays (PTAs) is to detect a stochastic background generated by the incoherent superposition of gravitational waves (GWs), in particular from the cosmic population of supermassive black hole binaries. Current stochastic background searches assume that pulsars in a PTA are separated from each other and the Earth by many GW wavelengths. As more millisecond pulsars are discovered and added to PTAs, some may be separated by only a few radiation wavelengths or less, resulting in correlated GW phase changes between close pulsars in the array. Here we investigate how PTA overlap reduction functions (ORFs), up to quadrupole order, are affected by these additional correlated phase changes, and how they are in turn affected by relaxing the assumption that all pulsars are equidistant from the solar system barycenter. 
We find that in the low frequency GW background limit of $f\sim10^{-9}$~Hz, and for pulsars at varying distances from the Earth, that these additional correlations only affect the ORFs by a few percent for pulsar pairs at large angular separations, as expected. 
However when nearby (order 100 pc) pulsars are separated by less than a few degrees, the correlated phase changes can introduce variations of a few tens of percent in the magnitude of the isotropic ORF, and much larger fractional differences in the anisotropic ORFs-- up to 188 in the $m=0$, $l=2$ ORF for equidistant pulsars separated by 3$\Degs$. In fact, the magnitude of most of the anisotropic ORFs is largest at small, but non-zero, pulsar separations. Finally, we write down a small angle approximation for the correlated phase changes which can easily be implemented in search pipelines, and for completeness, examine the behavior of the ORFs for pulsars which lie at a radiation wavelength from the Earth. 
\end{abstract}


\maketitle

\section{Introduction}
The direct detection of gravitational waves (GWs), predicted by Einstein's theory of gravity, will open a new window of observation on the Universe. The existence of GWs was confirmed via neutron star binary observations decades ago~\cite{HulseTaylor:1975, TaylorWeisberg:1982, Stairs:2006, KramerWex:2009}, however direct evidence of GWs remains elusive. The direct detection of GWs is possible with a Pulsar Timing Array (PTA)~\cite{HellingsDowns:1983, Sazhin:1978,
  Detweiler:1979, EstabrookWahlquist:1975}---
a type of GW detector which uses
one or more radio telescopes to regularly monitor a selection of
ultra-stable millisecond pulsars. The propagation time of radio waves
from each pulsar to the Earth is affected by the GW-induced space-time
perturbations along its path. The difference between the expected and
actual time-of-arrival of the radio pulses, called the timing
residual, carries information about the GWs which can be obtained by
correlating the residuals between pulsar pairs in the array.

PTAs are sensitive to gravitational radiation in the nanohertz frequency regime: 
the low-frequency limit is set by the inverse of the total observation time--  
typically 10 yrs-- yielding a lower bound of $1/(10$~yrs) $\sim 3\times 10^{-9}$~Hz. 
A few pulsars have been timed for approximately $30$ years, and are 
therefore sensitive to radiation of $ 10^{-9}$~Hz.
 The cadence of observation, typically a few weeks to months, gives
an upper frequency bound of $\sim 10^{-6}$~Hz. A promising class of
sources in this frequency band are supermassive black hole binary
(SMBHB) systems with masses in the range of $\sim 10^7 - 10^9$~$\Msun$
during their slow, adiabatic inspiral phase
~\cite{RajagopalRomani:1995, WyitheLoeb:2003, JaffeBacker:2003,
  SesanaVecchioColacino:2008, SesanaVecchioVolonteri:2008,
  WenEtAl:2011, SesanaMRASLett:2013}. GWs from other more speculative sources from the
early Universe, including cosmic strings~\cite{PshirkovTuntsov:2010,
  SanidasBattyeStappers:2012, KuroyanagiEtAl:2012} and relic GWs~\cite{Zhao:2011, Grischuk:1993}, are also expected to be found in this frequency
band. Searches of increasing sensitivity are currently ongoing in the
European PTA (EPTA) \cite{EPTA}, the Parkes PTA \cite{PPTA} and the
North American Nanohertz Gravitational Wave Observatory
(NANOGrav)~\cite{NANOGrav}, which together form the International PTA
(IPTA) \cite{IPTA}. 

In stochastic GW background searches, the expectoration value of the cross-correlated
timing residuals is proportional to the overlap reduction function (ORF)-- a dimensionless
function which quantifies the response of the pulsars to the
stochastic GW background, see e.g.~\cite{Finn:2009, ChamberlinSiemens:2012}. 
The ORF in turn depends on the frequency of the GW background and the geometry of the PTA, specifically the distance to each pulsar and the
angular separation of pulsar pairs. 
A standard normalization is usually applied to the ORFs which ensures that the maximum correlation between 
pulsar pairs, in an isotropic stochastic GW background, is $1.0$ for the ``auto-correlation"-- the correlation of a pulsar's timing residual with itself.

The observed timing residuals depend on a linear combination of the GW perturbation at the time when the
GW transits at the pulsar, the so-called ``pulsar term'', and then
when the GW passes the Earth, called the ``Earth term''. The Earth-term residual is then transformed to the solar system barycenter (SSB), which will be the origin of our geometry from hereon.
In stochastic GW background searches, the Earth-term contribution is correlated between all the
 pulsar pairs. If we assume that the inter-pulsar distance is large with respect to the radiation wavelength of the stochastic GW background, the pulsar term will add incoherently at the SSB due to large GW phase offsets between the pulsars, and average to zero when integrated over the sky. One may therefore simplify the ORF by considering the Earth term only. In this limit, called the ``short wavelength approximation'', the ORF is no longer dependent on pulsar distances and the frequency of the stochastic GW background.

 To date, all the stochastic GW
background searches use the short wavelength approximation, except when considering
the auto-correlation, which doubles the ORF~\cite{HellingsDowns:1983, Mingarelli:2013, TaylorGair:anisotropy}. 
However, as PTAs become more densely populated with millisecond pulsars, either by dedicated pulsar searches with
current radio telescopes, e.g. \cite{Barr:2013, Keith:2010,
  Stovall:2013}, or by searches performed by advanced radio telescopes
currently under development such as the Five Hundred Meter Aperture
Spherical Radio Telescope (FAST)~\cite{Nan:2008} and/or the
Square Kilometre Array (SKA)~\cite{Lazio:2013} to name but a few examples, pulsars in a PTA
may no longer lie many radiation wavelengths apart. This would be
especially true if millisecond pulsars in globular clusters,
e.g.~\cite{Manchester:1991, Wolszczan:1989, Lorimer:2008}, were
eventually added to PTAs.\footnote{Large uncertainties in the time of
  arrival of the radio pulses are introduced by the gravitational
  acceleration toward the core of the globular
  cluster, making them unsuitable 
  for current PTAs.}

Exploring the limits of the short wavelength approximation is the main theme of this paper.
If the inter-pulsar separation is no longer large with respect to the radiation wavelength of the stochastic GW background, the pulsar term may need to be included in the evaluation of the ORF to model the correlated phase changes between close pulsar pairs. To get a sense of how these correlated phase changes may impact stochastic GW background searches, we
examine how the pulsar term modifies the ORFs for close pulsar pairs bathed in a low-frequency stochastic GW background of $f\sim 10^{-9}$~Hz. Such a low-frequency background can be generated by SMBHBs in their slow, adiabatic inspiral phase, as well as by the cosmic
population of eccentric SMBHBs according to new studies by \cite{Sesana:2013, Ravi:2014}. This
type of low frequency GW background paired with nearby pulsars such as
J0437$-$4715 which is only $L\sim160$~pc from the Earth~\cite{ATNF} (or equivalently the SSB), yield a lower bound of 10 on the number
of radiation wavelengths, $fL$, which separate the SSB from the nearest pulsar in a PTA.
The following calculations can therefore be considered upper limits on how the correlated phase changes affect the ORFs. Although the low-frequency stochastic GW background is likely to be highly isotropic~\cite{Mingarelli:2013}, the anisotropic ORFs are included for completeness. Pulsars from the IPTA mock data challenge are used throughout to give  examples when additional phase terms should be included in the ORFs. 

The paper is organized as follows: in Sec \ref{s:signal} we provide a brief introduction to stochastic GW backgrounds which is largely taken from \cite{Mingarelli:2013}-- more details can be found in e.g. \cite{Anholm:2009, Mingarelli:2013,TaylorGair:anisotropy}. This is followed by an introduction to the PTA ORF in Sec \ref{s:ORF}, where we show how cross-correlated timing residuals can be used to search for stochastic GW backgrounds. In Sec \ref{s:distance} we illustrate how relaxing the assumption that all the pulsars in a PTA are at the same distance from the SSB affects the magnitude of the ORFs. This is done for the all the ORFs up to the quadrupole.
In Sec \ref{s:pulsarterm}, we approximate the pulsar term for neighboring pulsars separated by less than a few degrees. We show that this approximation captures the most important behavior of the pulsar term, and set a limit on when the pulsar term can be ignored. 
For completeness, we further investigate the behavior of the pulsar term when a pulsar is within one radiation wavelength from the Earth in Sec \ref{s:closepulsar}. Conclusions are presented in Sec \ref{sec:conclusion}.

We consider geometric units, and therefore set $c = G = 1$.
\section{Stochastic backgrounds}
\label{s:signal}
Let us consider a plane wave expansion for the metric perturbation $h_{ij}(t,\vec x)$ produced by a stochastic background:
\beq
h_{ij}(t,\vec x) = \sum_A \int_{-\infty}^\infty \!\!\!\!df\ \int_{S^2} \!\!\!
d\hat\Omega\ h_A(f,\hat \Omega)\ e^{i2\pi f(t-\hat \Omega\cdot\vec x)}\ 
e_{ij}^A(\hat \Omega)\ ,
\label{e:h_ab}
\eeq
where $f$ is the frequency of the GWs, the index $A = +\,,\times$ labels the two independent polarizations, the spatial indices are $ i,j=1,2,3 $, 
the integral is on the two-sphere $S^2$ and $d\hat\Omega=\sin\theta
d\theta d\phi$ where $\theta$ and $\phi$ are the usual polar and
azimuthal angles respectively, see Fig.\ \ref{fig:geometry}. 
The unit
vector $\hat\Omega$ identifies the propagation direction of a single
gravitational plane wave, that can be decomposed over the GW polarization tensors $e_{ij}^A(\hat\Omega)$ and the two independent polarization amplitudes,  $h_A(t, \hat \Omega)$ or equivalently $h_A(f,\hat \Omega)$~\cite{MTW, AllenRomano:1999}: 

\begin{subequations}
\begin{align}
h_{ij}(t,\hat\Omega) & = e_{ij}^+(\hat\Omega) h_+(t,\hat\Omega) + e_{ij}^{\times}(\hat\Omega)\, h_\times(t,\hat\Omega)\,,
\label{e:hab}
\\
h_{ij}(f,\hat\Omega) & = e_{ij}^+(\hat\Omega) h_+(f,\hat\Omega) + e_{ij}^{\times}(\hat\Omega)\, h_\times(f,\hat\Omega).
\label{e:hab_f}
\end{align}
\end{subequations}
The polarization tensors $e_{ij}^{A}(\hat\Omega)$ are uniquely defined once one specifies the wave principal axes described by the unit vectors $\hat m$ and $\hat n$: 
\begin{subequations}
\begin{align}
e_{ij}^+(\hat\Omega) &=  \hat m_i \hat m_j - \hat n_i \hat n_j\,,
\label{e:e+}
\\
e_{ij}^{\times}(\hat\Omega) &= \hat m_i \hat n_j + \hat n_i \hat m_j\,.
\label{e:ex}
\end{align}
\end{subequations}
For a stationary, Gaussian and unpolarized background, the polarization amplitudes satisfy the following statistical properties: 
\beq
\langle h^*_A(f,\hat{\Omega})  h_{A'}(f',\hat{\Omega}') \rangle = \delta^2(\hat{\Omega},\hat{\Omega}') \delta_{AA'} \delta(f - f') H(f) P(\hat\Omega)\,,
\label{e:hstat}
\eeq
where $\langle \cdot \rangle$ is the expectation value and $ \delta^2(\hat{\Omega},\hat{\Omega}')= \delta(\cos\theta-\cos\theta')\delta(\phi-\phi')$ is the covariant Dirac delta function on the two-sphere \cite{Finn2009}. This condition implies that the radiation from different directions are statistically independent. Moreover, we have factorized the power spectrum such that $P(f,\hat{\Omega})=H(f)P(\hat{\Omega})$, where the function $H(f)$ describes the spectral content of the radiation, and $P(\hat{\Omega})$ describes the angular distribution of the GW energy density on the sky. As in~\cite{Mingarelli:2013, AllenOttewill:1997}, we decompose the GW energy density on the basis of the spherical harmonic functions,

\beq
P(\hat{\Omega}) \equiv \sum_{lm} c_l^m Y_l^m(\hat{\Omega})\,,
\label{e:P_omega}
\eeq
where the sum is over $0 \le l < +\infty$, and $|m| \le l$. The anisotropy
coefficients, $c_l^m$, become additional search parameters in anisotropic
stochastic GW background searches (see e.g.~\cite{TaylorGair:anisotropy}), which
characterize the angular distribution of the background. Here the coefficients 
are assumed to be frequency independent, however, a future study is planned to introduce
a frequency-dependence in these coefficients.

\section{The overlap reduction function for Pulsar Timing Arrays}
\label{s:ORF}

The IPTA now shares data on over 40 millisecond pulsars which are regularly monitored by 8 radio telescopes: 5 in Europe, 2 in North American and 1 in Australia~\cite{IPTAmdc}. GWs affect the time of arrival of radio pulses at the SSB. Consider for example a pulsar with frequency $\nu_0$ whose location in the sky is described by the unit vector $\hat{p}$, at a distance $L$ from the SSB. 
\begin{figure}[hb]
		\centering
		\includegraphics[width=4in]{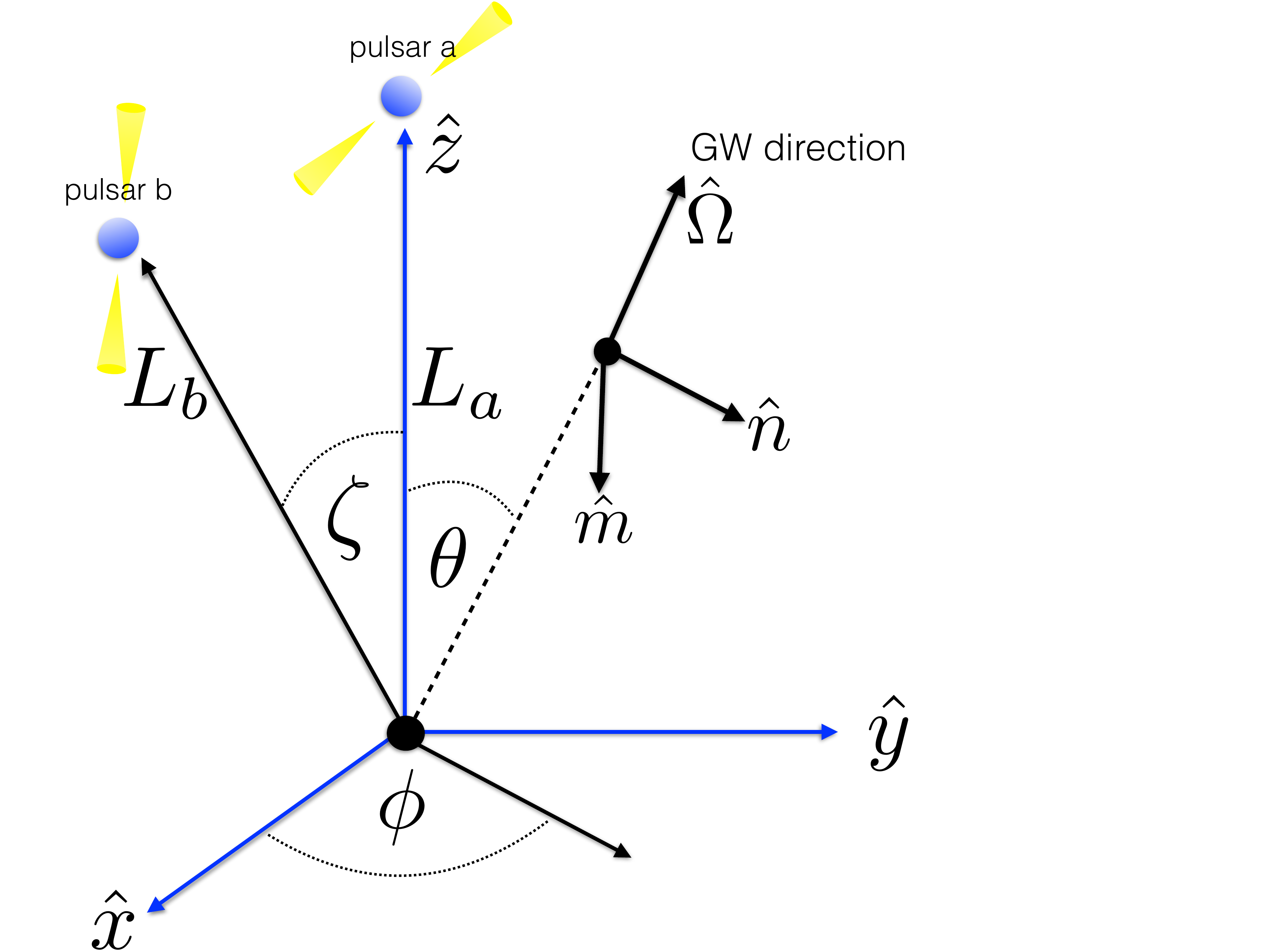}
		\caption[Typical pulsar timing array geometry]{The ``computational'' frame: pulsar $a$ is on the $z$-axis at a distance $L_a$ from the origin (solar system barycentre), pulsar $b$ is in the $x$-$z$ plane at a distance $L_b$ from the origin making an angle $\zeta$ with pulsar $a$. $\hat\Omega$ is the direction of GW propagation with principal axes $\hat m$ and $\hat n$ such that  $\hat m \times \hat n =\hat\Omega$. The polar and azimuthal angles are given by $\theta$ and $\phi$, respectively.}
		\label{fig:geometry}
	\end{figure}
A GW source in direction $-\hat\Omega$, see Fig \ref{fig:geometry},
generates a metric perturbation $h_{ij}(t,\hat{\Omega})$, affecting
frequency of the radio pulses, $\nu$, received at the radio telescope. This frequency shift
  is given by

\beq
z(t,\hat\Omega)  \equiv  \frac{\nu(t) - \nu_0}{\nu_0} = \frac{1}{2} \frac{\hat p^i\hat p^j}{1+\hat \Omega \cdot \hat p} \Delta h_{ij}(t,\hat{\Omega})\,,
\label{e:z}
\eeq
%
%%
% --------------  Include the figures for Gamma00 -------------------%
%%
\begin{figure*}[ht!]
     \begin{center}
        \subfigure[\, Isotropic overlap reduction function]{%
            \label{fig:00}
            \includegraphics[width=0.48\textwidth]{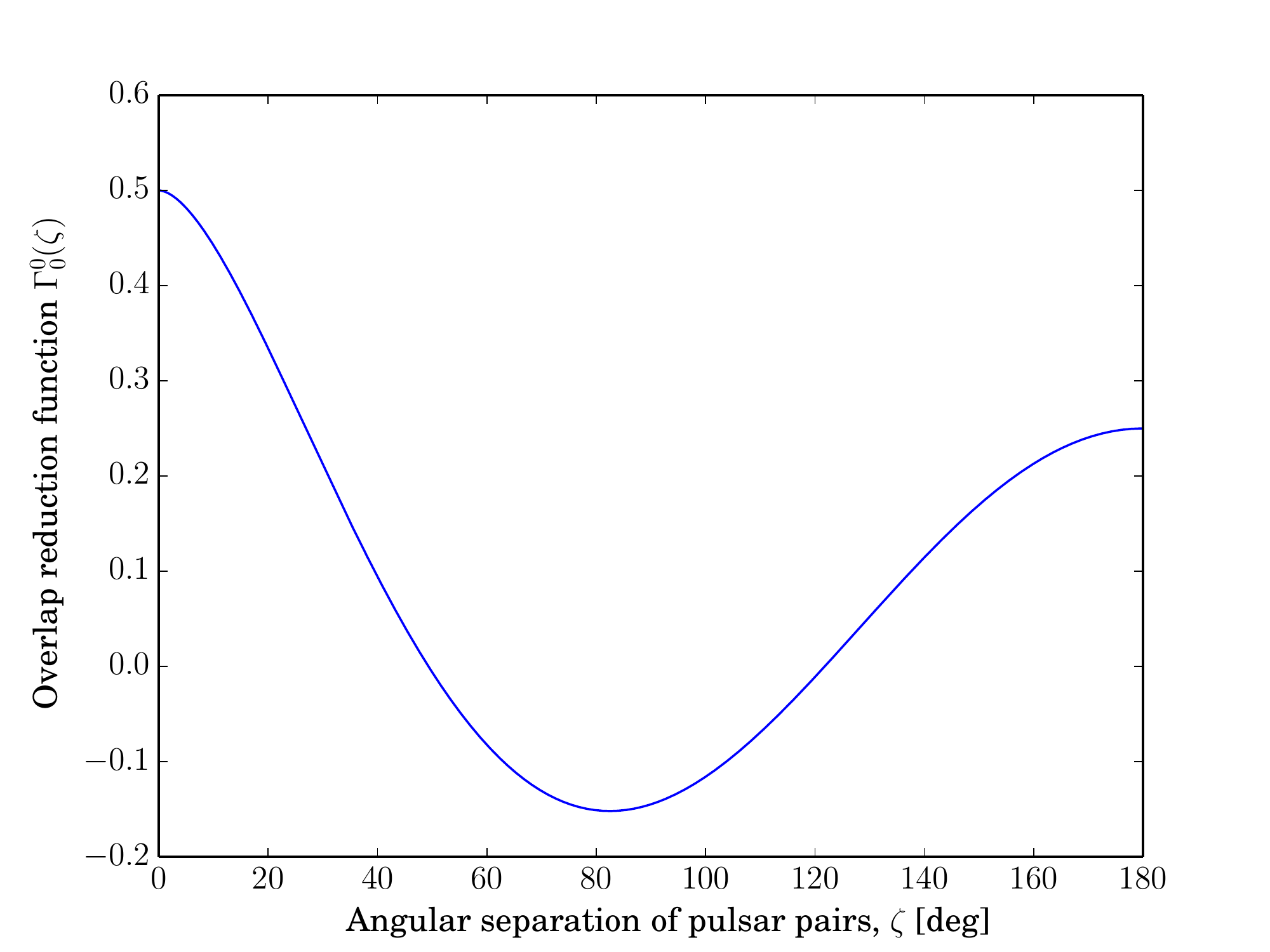}
        }%
	\subfigure[\, Dipole overlap reduction functions, $l=1$]
	{%
            \label{fig:m01}
            \includegraphics[width=0.48\textwidth]{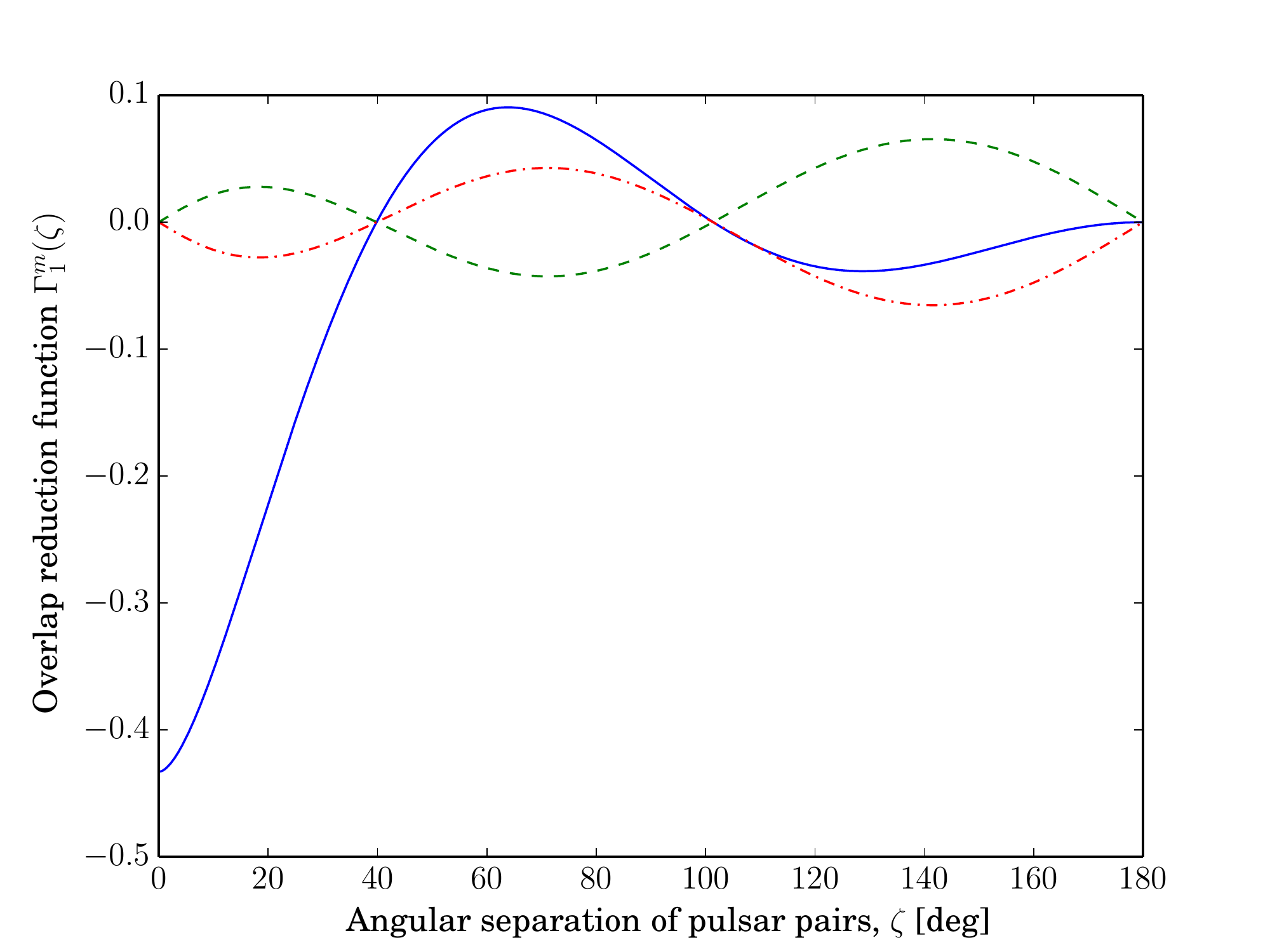}
        }\\ %  ------- End of the first row ----------------------%
        \subfigure[\, Quadrupole overlap reduction functions, $l=2$]{%
            \label{fig:m012}
            \includegraphics[width=0.49\textwidth]{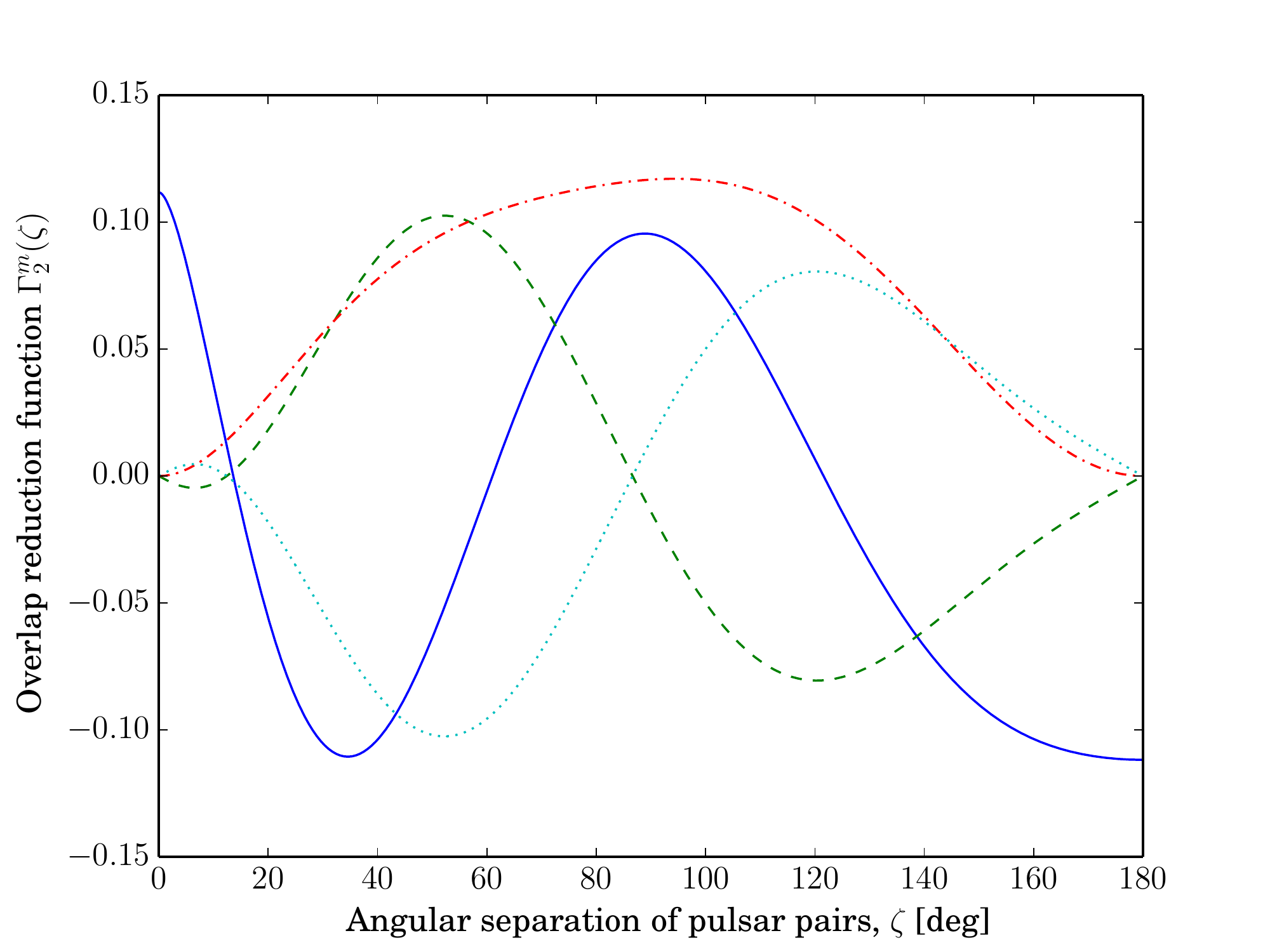}
        }%
    \end{center}
  
    \caption{%
The Earth-term-only overlap reduction functions for isotropic, Fig~\ref{fig:00}; dipole, Fig~\ref{fig:m01}; and quadrupole Fig~\ref{fig:m012} GW energy density distributions~\cite{Mingarelli:2013}. In the chosen reference frame, $\Gamma^{-m}_l(\zeta)=(-1)^m\Gamma^m_l(\zeta)$. The legends are as follows: in Fig~\ref{fig:m01}, the $m=0$ curve is the solid (blue) curve, $m=1$ is the dashed (green) curve and $m=-1$ is the dashed-dot (red) curve. In Fig \ref{fig:m012}, $m=0$ is the solid (blue) curve, $m=1$ is the dashed (green) curve, $m=-1$ is the dotted (cyan) curve and $m=\pm 2$ is the dashed-dot (red) curve. Features of these ORFs are discussed in Sec \ref{sec:ETfeatures}.}%
 \label{fig:original_curves}
\end{figure*}
where
\beq
\Delta h_{ij}(t,\hat\Omega) \equiv h_{ij}(t,\hat{\Omega}) - h_{ij}(t_\mathrm{p},\hat{\Omega}) 
\label{e:deltah}
\eeq
is the difference between the metric perturbation at the SSB $h_{ij}(t,\hat{\Omega})$, the {Earth term}, with coordinates $(t,\vec{x})$, and at the pulsar $h_{ij}(t_\mathrm{p},\hat{\Omega})$, the { pulsar term}, with coordinates $(t_\mathrm{p},\vec{x}_p)$. We consider a frame in which 
\begin{subequations}
\begin{align}
t_p & = t_e - L = t - L \quad \quad \vec{x}_p = L \hat p \,,
\\
t_e & = t \quad \quad \vec{x}_e = 0\,,
\end{align}
\end{subequations}
where the indices ``e'' and ``p'' refer to the Earth (SSB) and the pulsar. 
In this frame we can therefore write Eq.~\eqref{e:deltah} using Eq.~\eqref{e:hab_f} 
\bea
\Delta h_{ij}(t,\hat \Omega) = &&
\sum_A \int_{-\infty}^\infty df e_{ij}^A(\hat \Omega) \ h_A(f,\hat \Omega)\
\!e^{i 2 \pi f t}\ \!\!
\nonumber\\
&& 
\times \left[1 - e^{-i 2 \pi f L (1+ \hat \Omega \cdot \hat p)}\right]\, . 
\label{e:deltah1}
\eea
The fractional frequency shift, $z(t)$, produced by a stochastic background is simply given by integrating Eq.~(\ref{e:z}) over all directions. Using Eq.~(\ref{e:deltah1}), we obtain:
\bea 
\label{eq:freqShift}
z(t)  &=& \int d\hat\Omega\, z(t,\hat\Omega)
\nonumber \\
& = & \sum_A \int_{-\infty}^\infty df \int_{S^2}d\hat\Omega F^A(\hat\Omega) h_A(f,\hat \Omega)\nonumber\\
&&\times e^{i 2 \pi f t}\ \left[1 -  e^{-i 2 \pi f L (1+ \hat \Omega \cdot \hat p)} \right],
\eea
where $F^A(\hat\Omega)$ are the antenna beam patterns for each polarization $A$, defined as
\beq
F^A(\hat\Omega) =\left[\frac{1}{2} \frac{\hat p^i\hat p^j}{1+\hat \Omega \cdot \hat p} \ e_{ij}^A(\hat \Omega)\right].
\label{e:F+Fx}
\eeq
The quantity that is actually observed is the time-residual $r(t)$, which is simply the time integral of Eq.~(\ref{eq:freqShift}):
\beq
r(t) = \int^t dt' z(t')\,. \\
\label{e:r}
\eeq

Searches for a stochastic GW background rely on looking for correlations induced by GWs in the timing residuals from pulsar pairs. The expected value of the correlation between a residual from pulsar $a$ at time $t_j$, with that from a different pulsar, say pulsar $b$ at time $t_k$, depends on terms of the form:

\beq
\langle r_a^*(t_j) r_b(t_k) \rangle = 
\left\langle \int^{t_j} dt' \int^{t_k} dt'' z_a^*(t') z_b(t'') \right\rangle \, ,
\eeq

\begin{widetext}
\beq
\langle r_a^*(t_j) r_b(t_k) \rangle 
=  \int^{t_j} dt' \int^{t_k} dt'' \int_{-\infty}^{+\infty} df e^{- i 2\pi f(t' - t'')} H(f)\, ^{(ab)}\Gamma(f),
\label{e:corr}
\eeq
where $H(f)$ contains the information of the spectrum of radiation. In analogy with~\cite{AllenRomano:1999, Mingarelli:2013, TaylorGair:anisotropy}, we define the quantity above that depends on the relative location of the pulsars in the PTA, and the angular distribution of the GW energy density as the \textit{overlap reduction function}, Fig \ref{fig:original_curves}:

\beq\label{eq:HDsol}
^{(ab)}\Gamma^m_l(fL, \zeta) \equiv \int d\hat\Omega \, Y^m_l(\theta,\phi) \kappa_{ab}(f,\hat\Omega) \left[\sum_A F_a^A(\hat\Omega) F_b^A(\hat\Omega)\right],
\eeq
where

\end{widetext}
\beq
\kappa_{ab}(f,\hat\Omega) \equiv \left[ 1 - e^{i 2 \pi f L_a (1+ \hat \Omega \cdot \hat p_a)} \right] \left[ 1 - e^{-i 2 \pi f L_b (1+ \hat \Omega \cdot \hat p_b)}\right].
\label{eq:kappa}
\eeq
For these investigations we use a particular reference frame,
called the ``computational frame'',  where one pulsar is placed along the
$z$-axis and the second in the $x$-$z$ plane, and the angle between the pulsars is
$\zeta$, as seen in Fig \ref{fig:geometry}. Specifically, we write

\begin{subequations}
	\label{e:coord}
	\begin{align}
	\hat p_a		&=(0,0,1), \\
	\hat p_b	&=(\sin\zeta,0,\cos\zeta),\\
	\hat \Omega 	&=(\sin\theta\cos\phi,\sin\theta\sin\phi,\cos\theta),\\
	\hat m		&=(\sin\phi,-\cos\phi,0),\\
	\hat n		&=(\cos\theta\cos\phi,\cos\theta\sin\phi,-\sin\theta),
	\end{align}
\end{subequations}
where $\hat p_a$ and $\hat p_b$ are the unit vectors pointing to
pulsars $a$ and $b$, respectively, $\hat \Omega$ is the direction of
GW propagation and $\hat m$ and $\hat n$ are the GW principal axes,
see Eqs. \eqref{e:e+} and \eqref{e:ex} and Fig \ref{fig:geometry}. This is indeed a convenient
choice of geometry, as in this reference frame $F_a^\times = 0$ by Eq. \eqref{e:F+Fx}.

We now consider the behavior of the function $\kappa_{ab}(f,\hat\Omega)$, defined in Eq.~(\ref{eq:kappa}) and present in Eq.~(\ref{eq:HDsol}), which introduces the frequency and distance-dependence of the ORFs. Indeed, one can think of  $\kappa_{ab}(f,\hat\Omega)$ as the term which encodes the information about both the pulsar terms. Assuming $L_a=L_b=L$, the typical scale of $\kappa_{ab}(f,\hat\Omega)$ for the current pulsar population and PTA sensitivity is:
\beq
f L (1+ \hat \Omega \cdot \hat p) = 10^3 \left(\frac{f}{10^{-8}\,\mathrm{Hz}}\right)\!\! \left(\frac{L}{1\,\mathrm{kpc}}\right)\,
(1+ \hat \Omega \cdot \hat p).
\label{e:fL}
\eeq
The quantity $fL$ has a physical interpretation as the number of radiations wavelengths which, in this case, separate the pulsar from the SSB. It is a key physical quantity which will be used throughout the rest of this paper. 

When one computes the integral in Eq.~(\ref{eq:HDsol}) for large $fL$, the frequency dependent contributions to the integral rapidly average to zero as the angle between the pulsar pairs increases \cite{Anholm:2009, Mingarelli:2013}. Therefore,  Eq.~(\ref{eq:HDsol}) is well approximated by 
\beq
^{(ab)}\Gamma_l^m(\zeta) \simeq (1 + \delta_{ab}) \int d\hat\Omega\, Y^m_l(\theta,\phi) \left[\sum_A F_a^A(\hat\Omega) F_b^A(\hat\Omega)\right] \,,
\label{e:abGammalm1}
\eeq
where $\delta_{ab}$ is the Kronecker delta. Note that the above is an Earth-term only expression, except for the auto-correlation (when pulsar~$a=b$), and is therefore denoted by $^{(ab)}\Gamma_l^m(\zeta)$, whereas the full ORF is denoted by $^{(ab)}\Gamma_l^m(fL,\zeta)$.

Pulsars such as  J0437$-$4715, J1856$-$3754 and J2144$-$3933 lie at $160$~pc from the Earth (equivalently the SSB)~\cite{ATNF}, and such nearby pulsars in a low frequency GW background of 1~nHz, generated for example by eccentric SMBHBs~\cite{Sesana:2013, Ravi:2014} or simply SMBHBs in their slow adiabatic inspiral phase,  would be at $\sim 10$ radiation wavelengths from the SSB:
\beq
f L (1+ \hat \Omega \cdot \hat p) = 10 \left(\frac{f}{10^{-9}\,\mathrm{Hz}}\right)\!\! \left(\frac{L}{100\,\mathrm{pc}}\right)\,
(1+ \hat \Omega \cdot \hat p).
\label{e:fL_small}
\eeq
It is not clear that Eq. \eqref{e:abGammalm1} holds if the distances to the pulsars are small (order of 100 pc) and/or if the pulsar distances are allowed to vary ($L_a\neq L_b$). Moreover, it is not obvious that the $(1+\delta_{ab})$ approximation of $\kappa_{ab}(f,\hat\Omega)$ holds in Eq. \eqref{e:abGammalm1} if the pulsars are within a few radiation wavelengths, $fL$, of each other. These assumptions are verified in Sec \ref{s:distance}. 

\subsection{Features of the Earth term only ORFs}
\label{sec:ETfeatures}
The response of a PTA to a stochastic GW background depends on the position of the pulsars in the PTA and the distribution of the GW energy density on the sky. This response is captured in the ORFs, see Eq. \eqref{eq:HDsol}, and here we examine some of their features, in the limit that $fL\gg1$. The pulsar term is examined in detail in Sec \ref{s:distance}, and the description of its effect on the ORF is fully explored there. 

We note that for an isotropic stochastic GW background, the detector response for $\zeta=0$ is twice that of $\zeta=\pi$, see Fig \ref{fig:00}. Considering the response to an incoming GW at some angle $\theta$ may help one to understand this observation. 
If $\zeta=0$, which is the case for coincident and co-aligned pulsars (i.e. $a=b$), the antenna beam pattern, Eq. \eqref{e:F+Fx}, is given by
\beq
\label{eq:Faplus}
F^+_aF^+_{a=b}=\left[-\half\frac{\sin^2\theta}{1+\cos\theta}\right]^2=\quarter (1-\cos\theta)^2,
\eeq
where the numerator has been computed from $p^ip^je^+_{ij}=-\sin^2\theta$ for pulsar $a$ on the $z$-axis. Note that in this particular geometry, there is no $\phi$ dependence. Integrating this response over $d\hat \Omega=\sin\theta d\theta$ gives
\beq
\quarter \int_0^\pi d\theta \sin\theta(1-\cos\theta)^2=\frac{2}{3}.
\eeq
When $\zeta=\pi$, the antenna beam pattern is given by
\beq
F^+_aF^+_b=\half(1-\cos\theta)\half[1-\cos(\pi-\theta)]=\quarter \sin^2\theta\, ,
\eeq
and integrating over $\theta$ yields
\beq
\quarter \int_0^\pi d\theta \sin\theta(\sin^2\theta)=\frac{1}{3}.
\eeq
It is therefore clear that particular geometries are more (or less) sensitive to stochastic background signals. Note that this is an Earth-term-only argument, and does not take into account the pulsar term which doubles the ORF at $\zeta=0$ (the auto-correlation). 

More generally, features of the ORFs can be explained in terms of the alignment of the GW direction, $\hat\Omega$, and the position of the pulsar, $\hat p$, see Fig \ref{fig:geometry}. The product $\hat\Omega\cdot\hat p$ enters into the ORF via the antenna beam patterns given in Eq. \eqref{e:F+Fx}, where $F^+_{a,b}\propto (1+\hat\Omega\cdot\hat p)^{-1}$, and $\kappa_{ab}(f,\hat\Omega)$, Eq.~\eqref{eq:kappa}.  When $\hat\Omega$ is parallel or antiparallel to $\hat p$, $\hat\Omega\cdot\hat p=\pm1$. These cases both yield zero detector response, for reasons described below.

When $\hat \Omega\cdot\hat p=-1$, the photons emitted from the pulsar surf the GWs, and there is no frequency shift in the photons emitted from the pulsar. This effect can be understood by considering the metric perturbation including the pulsar term: since the signal at the SSB is the same as the signal at the pulsar, $\Delta h_{ij}(t,\hat\Omega)=0$ by Eq. \eqref{e:deltah}. Note however that there appears to be a divergence in the antenna beam pattern caused by zero division for the antiparallel case. In fact, since the pulsar position and the direction of GW propagation are antiparallel, $\theta=\pi$, and therefore by Eq.~\eqref{e:F+Fx}, $F^+_a=+1$ and is not singular. Recall that ORF is integrated over the whole sky and this is just one piece of the integration. 

When  $\hat\Omega\cdot\hat p=+1$, it is clear that $F^+_a=0$ by Eq.~\eqref{e:F+Fx} (equivalently Eq.~\eqref{eq:Faplus} if one considers only $F^+_a$), since $\theta=0$. Note that in this case the photons from the pulsar travel over the maximum number of radiation wavelengths resulting in a significant amount of redshifting, or ``stretching and squashing'', cf. Fig~\ref{fig:geometry}. The additional phases introduced by the GW then largely cancel out, limiting the detector response. 

Next we describe some features of the $l=1$, $m=0$ dipole ORF, Fig \ref{fig:m01} (solid blue line). Here the auto-correlation is negative, which may be counter-intuitive to the reader. This feature can be understood by considering the dipole spherical harmonic, $Y^0_1(\theta,\phi)$, which is a peanut shape with its maximum positive region aligned with $\theta=0$, and maximum negative region aligned on $\theta=\pi$. Here the $\theta$-axis is aligned with the $z$-axis, where we typically place pulsar $a$, see Fig~\ref{fig:geometry}. Since the direction of the source in this reference frame is actually $-\hat\Omega$, the GW energy density described by the dipole is inverted: hence the negative region aligns with $\hat{p}_a$ resulting in a negative auto-correlation. These effects are also explored for the quadrupole $Y^0_2(\theta,\phi)$ harmonic in Appendix A.
\section{Correlated phase changes from the pulsar term}
\label{s:distance}
\begin{figure}[b]
		\includegraphics[scale=0.25]{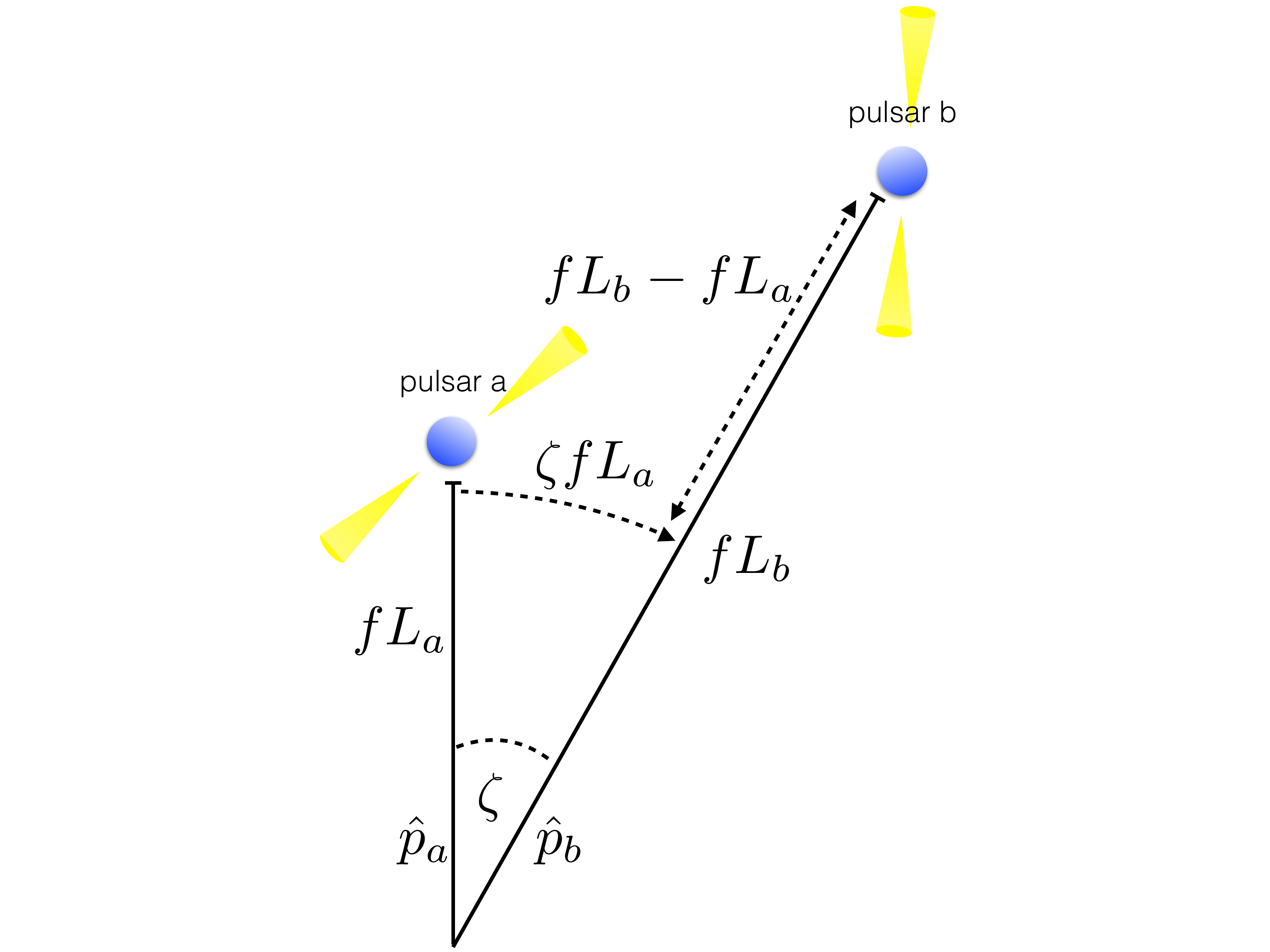}
		\caption{Geometry of pulsar in the ``strong
                  pulsar term regime''. Here $L_{x}$ is the distance to pulsar $x$
                  from the solar system barycentre (SSB) and $\zeta$ is the angular separation of
                  the pulsars. The dimensionless product $fL_x$
                  is the number of gravitational radiation wavelengths from the
                 SSB to each pulsar.
                 The geometry indicates two
                  possible movements: pulsar $b$ is moved
                  azimuthally by $\zeta fL_a$
                  radiation wavelengths from $a$, or $b$ is
                  moved radially away from the origin by $fL_b-fL_a$.} 
		\label{fig:pulsar_cartoon}
\end{figure}

The Earth-term ORF, Eq.~\eqref{e:abGammalm1}, is a good approximation to the full ORF  when 
$fL\gg1$, cf. Eq. \eqref{e:fL}. However, it is not clear that this form of the ORF can be
used in a selection of cases where the pulsars no longer lie at many radiation wavelengths
from the SSB, and/or from each other.

Here we explore how relaxing the assumption that all pulsars in a PTA
are at the same distance from the SSB
i.e. $L_a=L_b$ (see Figs \ref{fig:geometry}, \ref{fig:pulsar_cartoon}), affects the ORFs for nearby pulsars in the current low
frequency limit of the stochastic GW background. Since we have a
concrete lower bound of $fL=10$, we fix the dimensionless product
$fL_a=10$ and vary $fL_b$ from $10$ to $14$. Larger values of $fL_b$
were computed, up to $fL_b=20$ for all the ORFs, however the
oscillations converged to zero increasingly rapidly as $fL$
increased.

\begin{figure*}[p]
     \begin{center}
        \subfigure[\, Magnitude of  the isotropic overlap reduction function.]{%
            \label{fig:mag00}
            \includegraphics[width=0.48\textwidth]{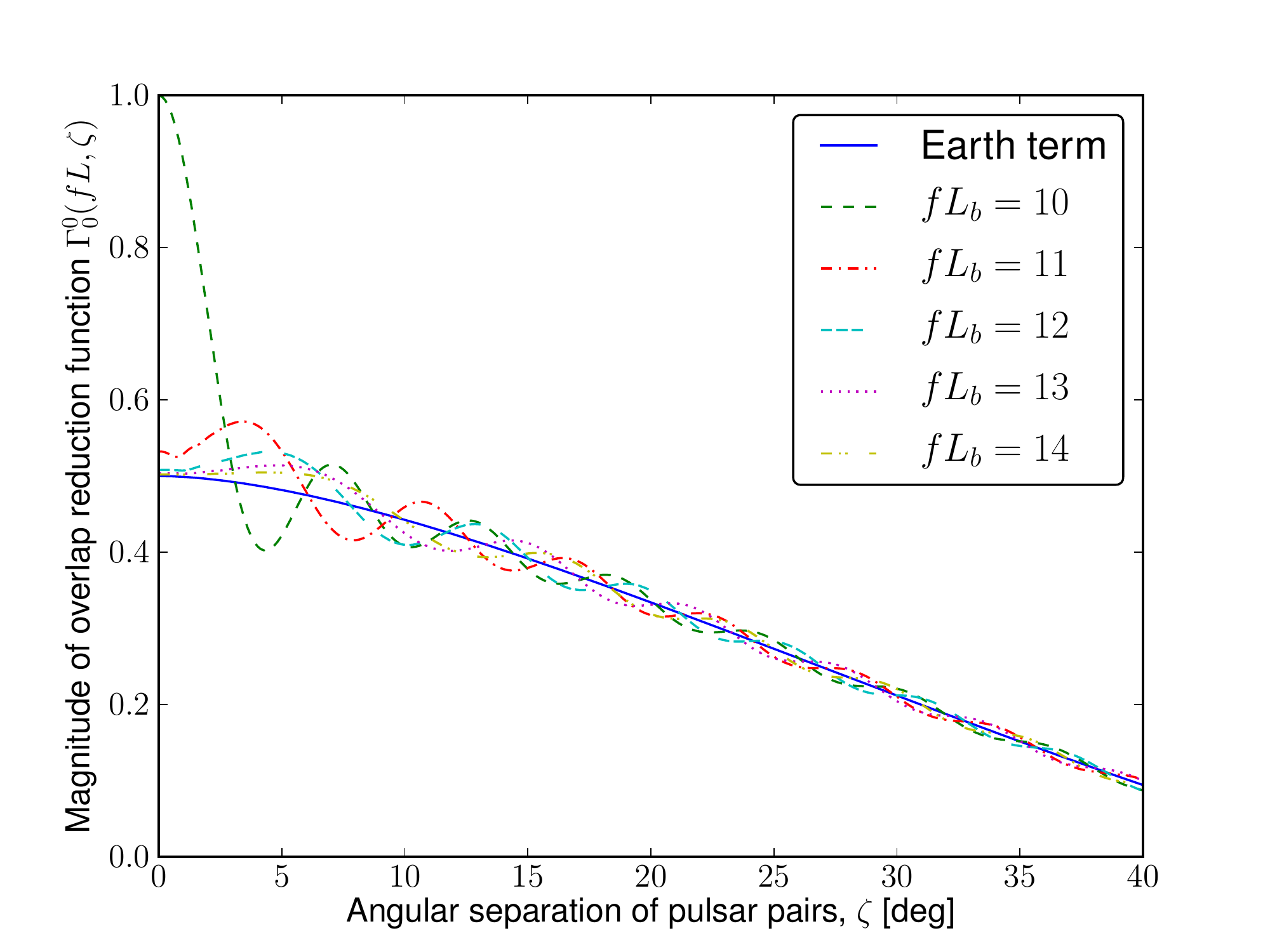}
        }%
	\subfigure[%
        \, Strong pulsar term regime for $^{(ab)}\Gamma^0_0(fL,\zeta)$.        
        ]{%
            \label{fig:contour00}
            \includegraphics[width=0.48\textwidth]{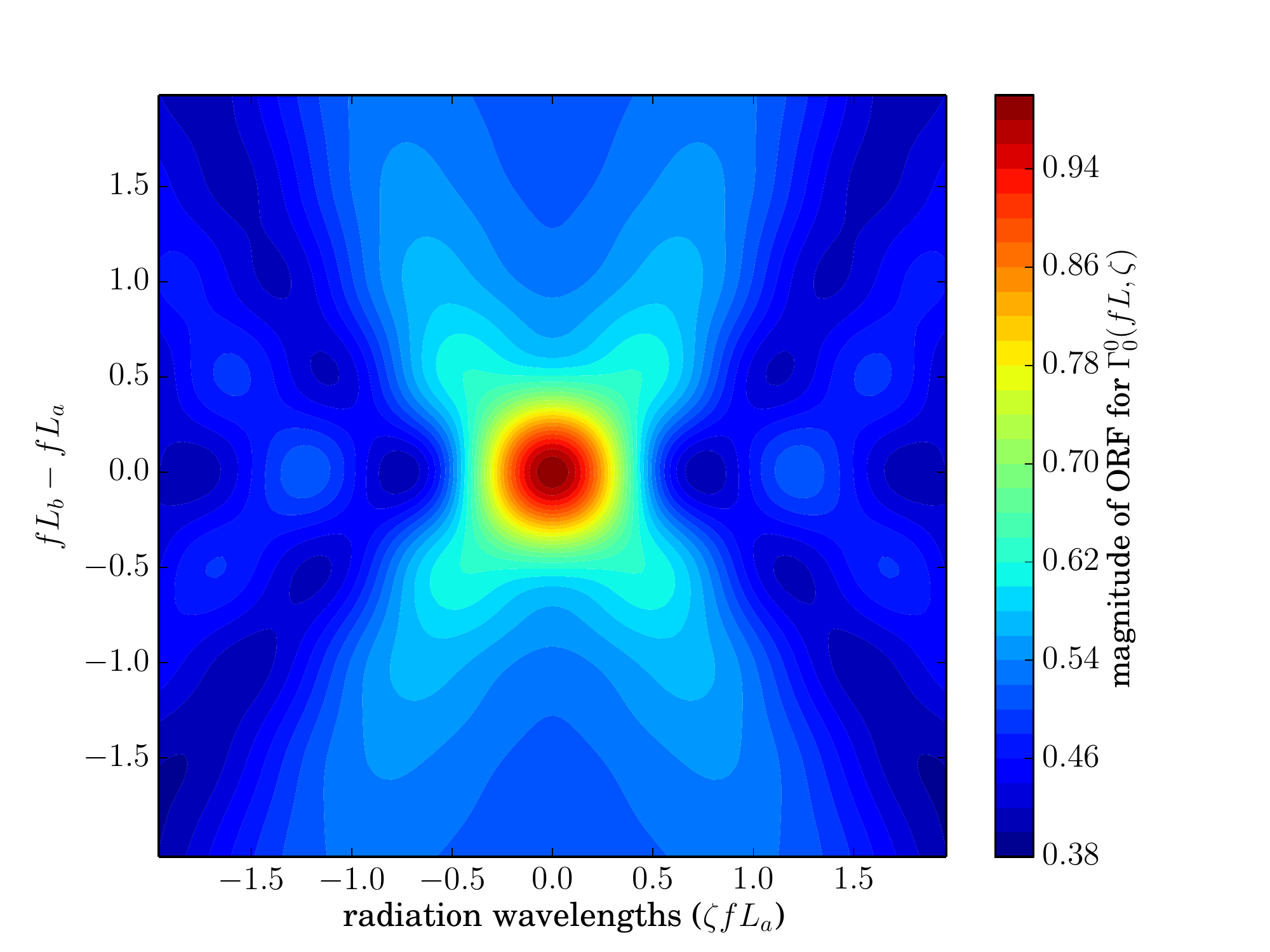}
        }\\ %  ------- End of the first row ----------------------%
        \subfigure[\, Magnitude of dipole $^{(ab)}\Gamma^0_1(fL,\zeta)$ ORF]{%
            \label{fig:mag01}
            \includegraphics[width=0.49\textwidth]{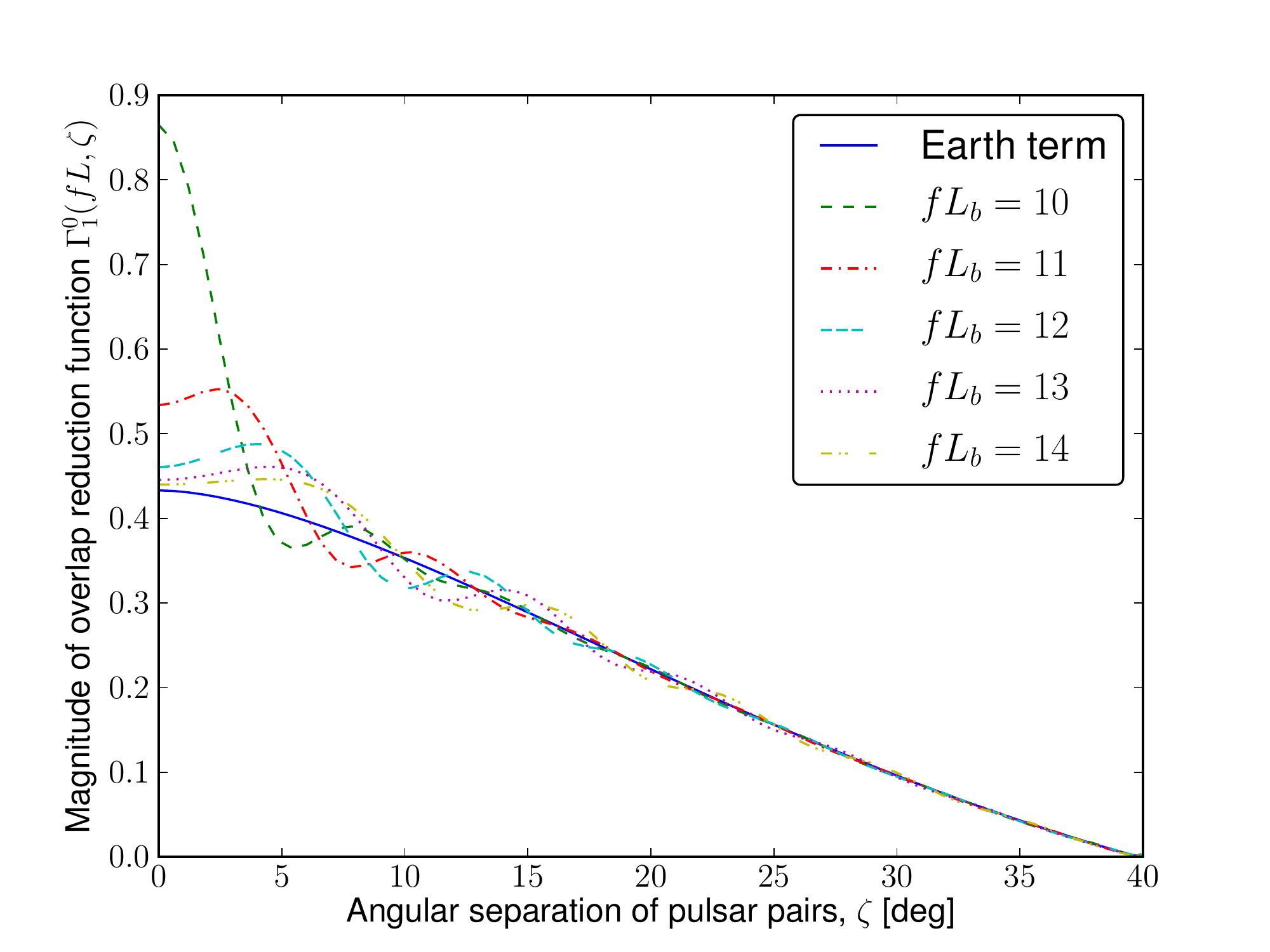}
        }%
	 \subfigure[%
        \, Strong pulsar term regime for $^{(ab)}\Gamma^0_1(fL,\zeta)$       ]{%
            \label{fig:contour01}
            \includegraphics[width=0.49\textwidth]{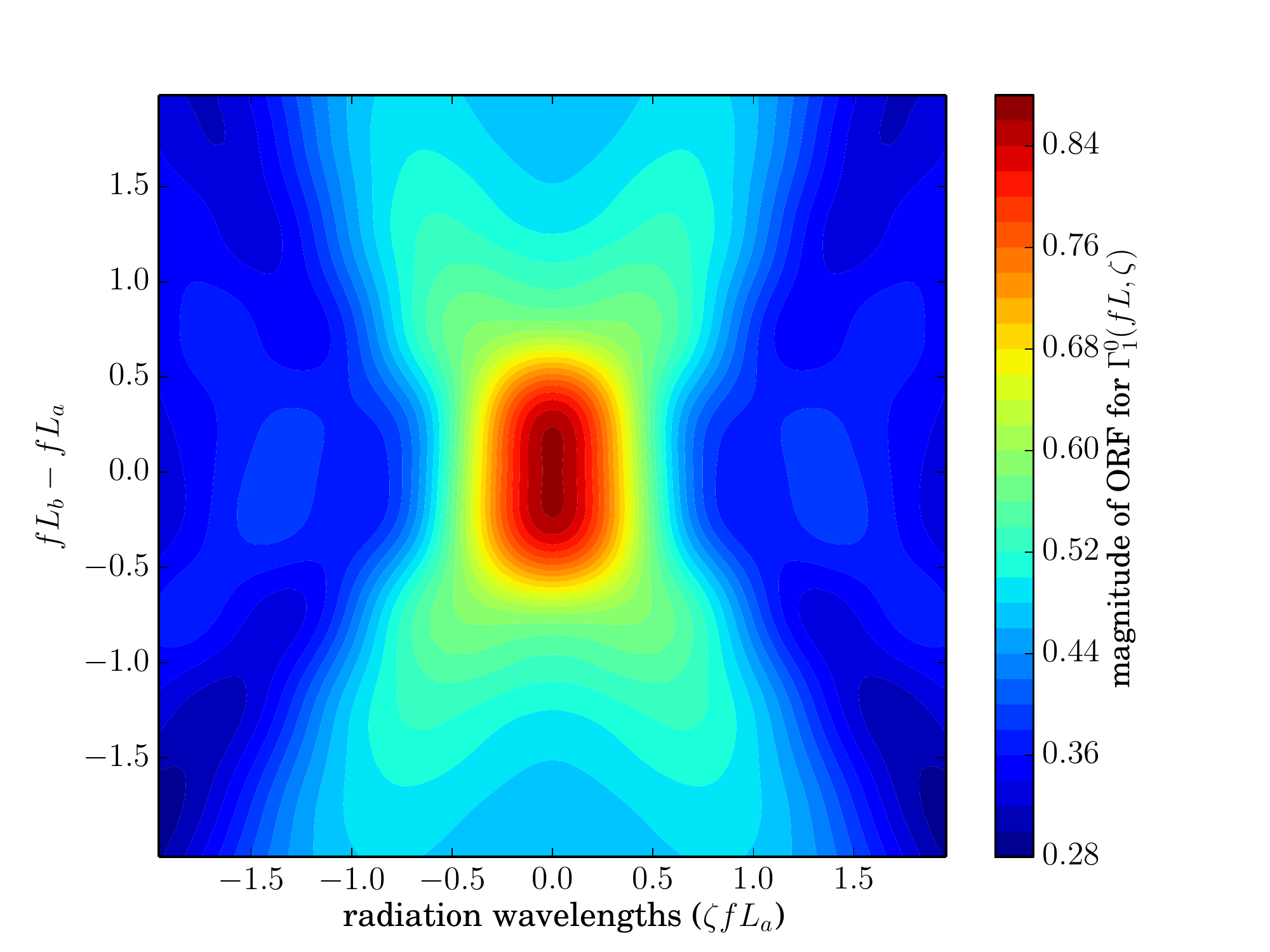}
        }%\\

 \subfigure[%
        \, Magnitude of the dipole $^{(ab)}\Gamma^1_1(fL,\zeta)$ ORF ]{%
            \label{fig:mag11}
            \includegraphics[width=0.49\textwidth]{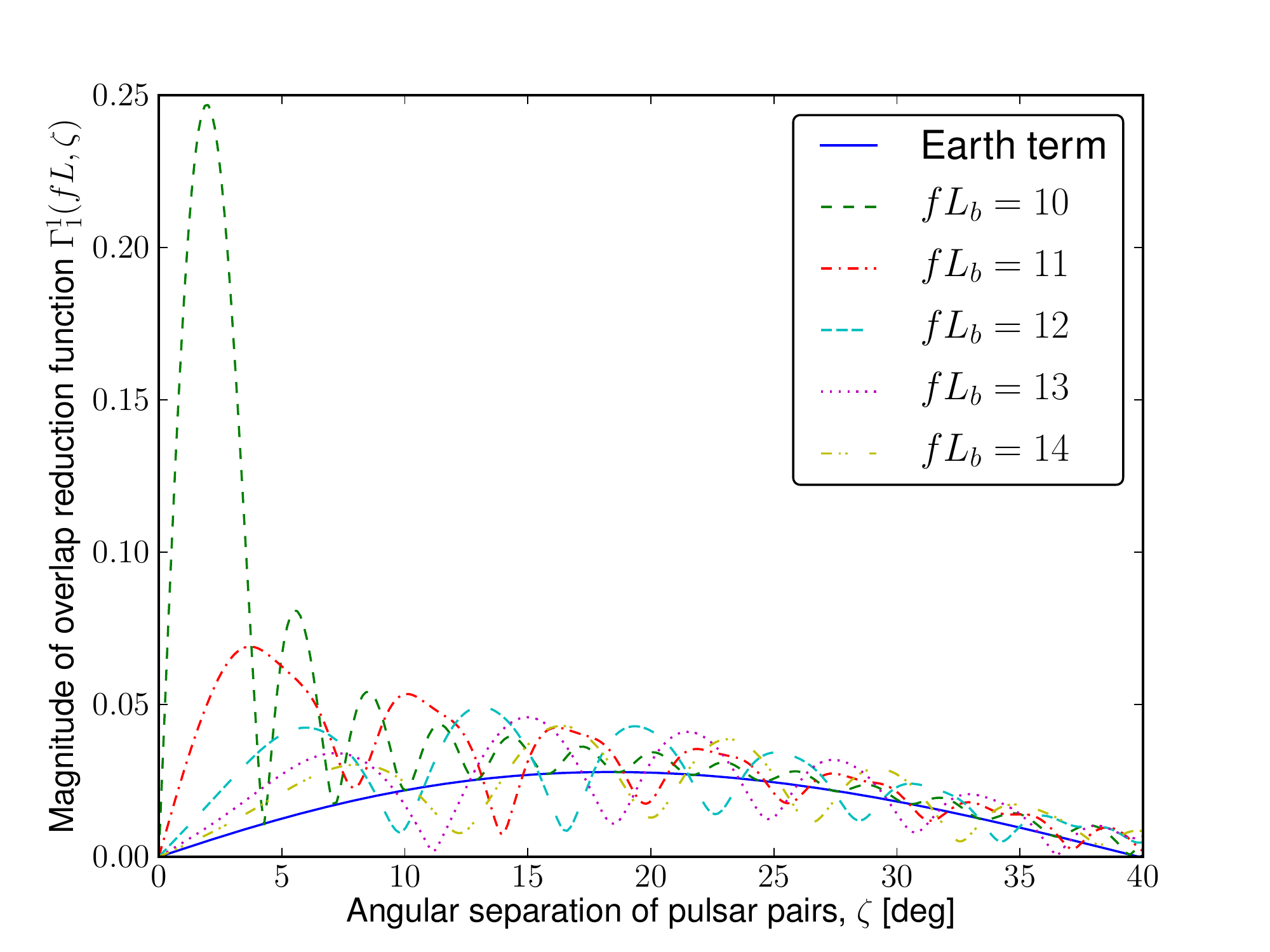}
        }%
 \subfigure[%
        \, Strong pulsar term regime for $^{(ab)}\Gamma^1_1(fL,\zeta)$]{%
            \label{fig:contour11}
            \includegraphics[width=0.49\textwidth]{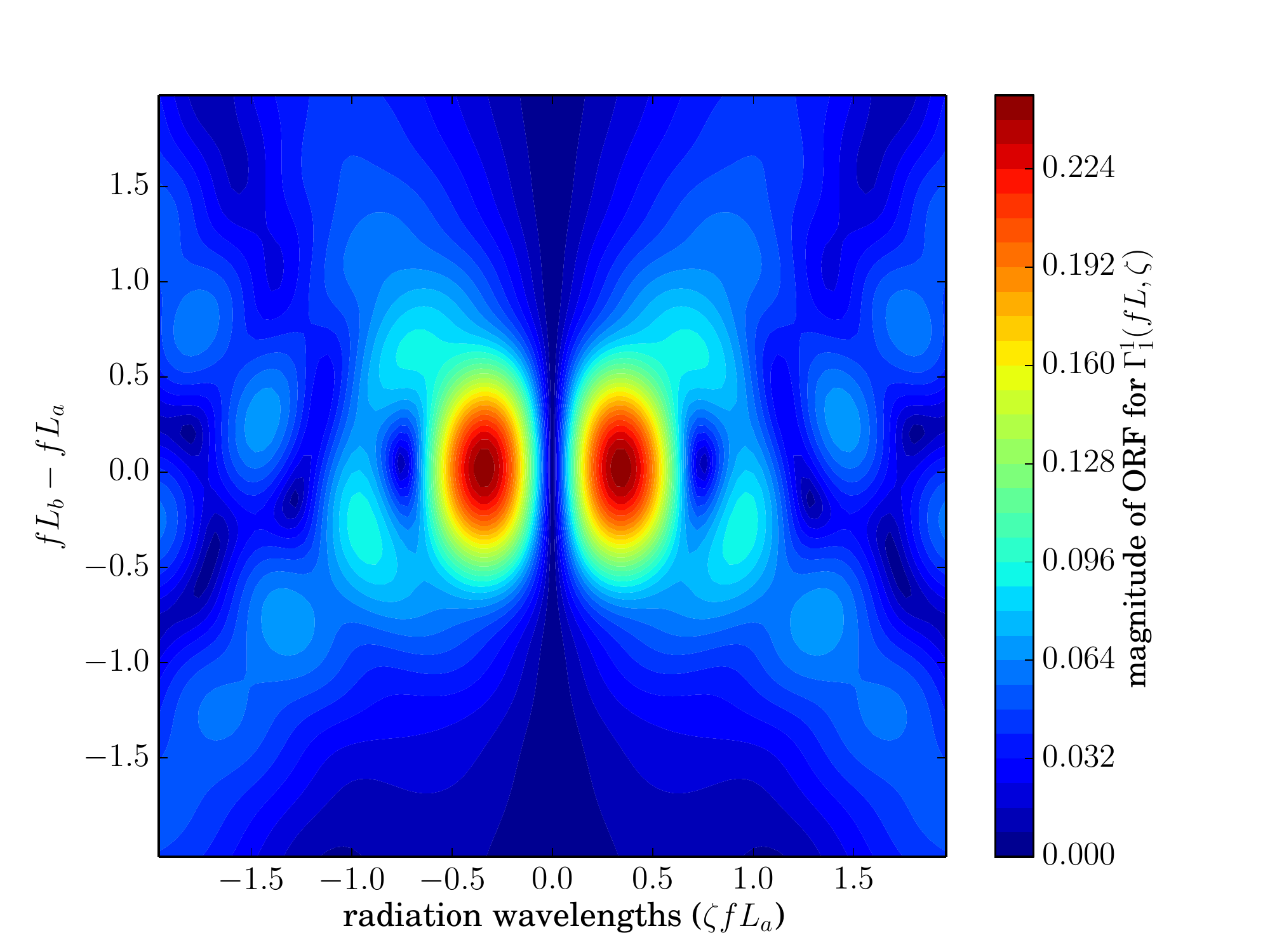}
        }%
    \end{center}
  
    \caption{%
The effect of pulsar distance variations on the magnitude of the isotropic and dipole overlap reduction functions, with $fL_a=10$ fixed. Panels on the left hand side are truncated at 40 degrees, as pulsar term oscillations rapidly converge to zero.}%
 \label{fig:00-11}
\end{figure*}

\begin{figure*}[p]
     \begin{center}
        \subfigure[\, Magnitude of the quadrupole $^{(ab)}\Gamma^0_2(fL,\zeta)$ ORF.]
	{%
            \label{fig:mag02}
           \includegraphics[width=0.48\textwidth]{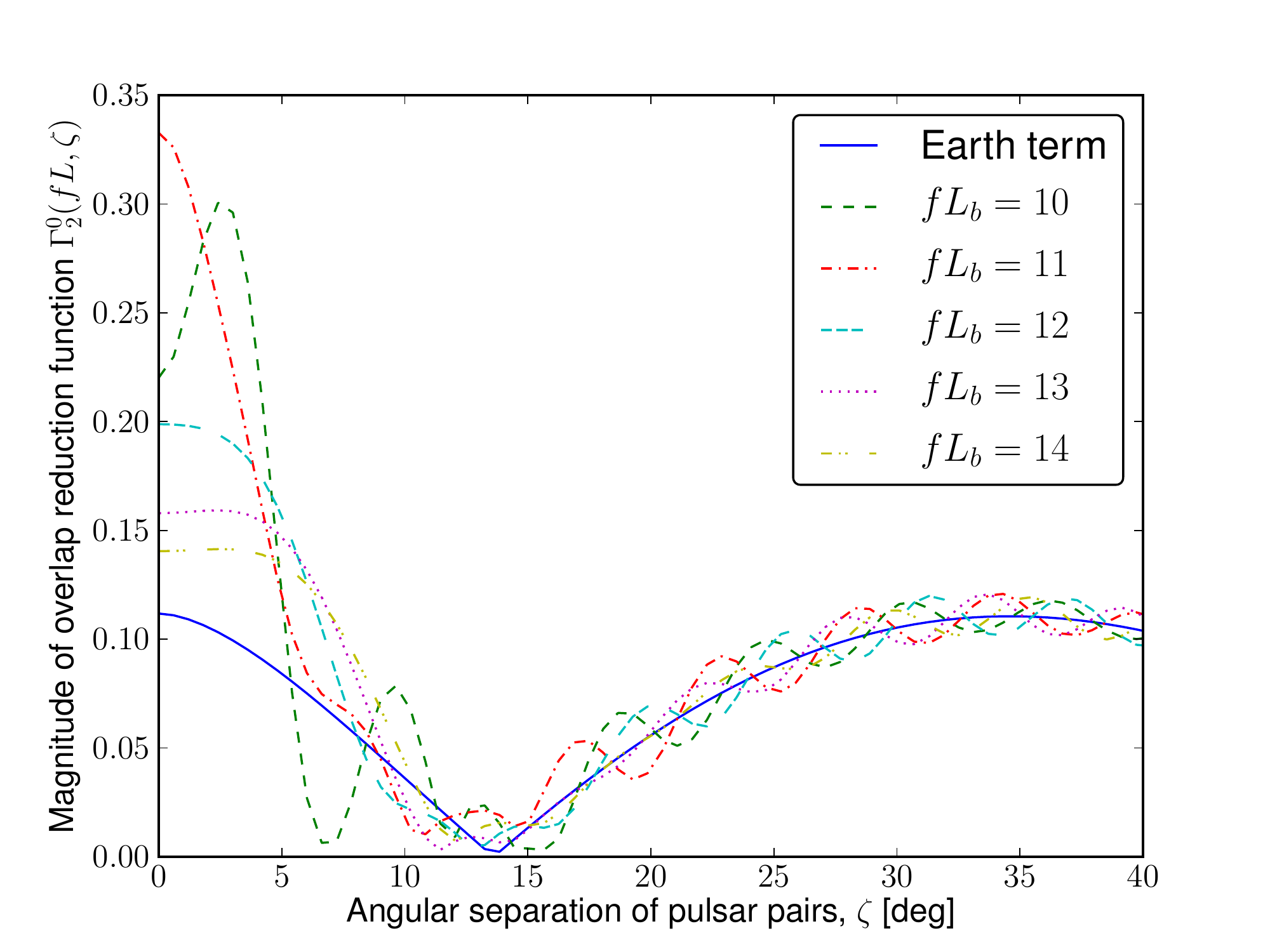}
        }
        \subfigure[%
        \,Strong pulsar term regime for  $^{(ab)}\Gamma^0_2(fL,\zeta)$]{%
            \label{fig:contour02}
            \includegraphics[width=0.48\textwidth]{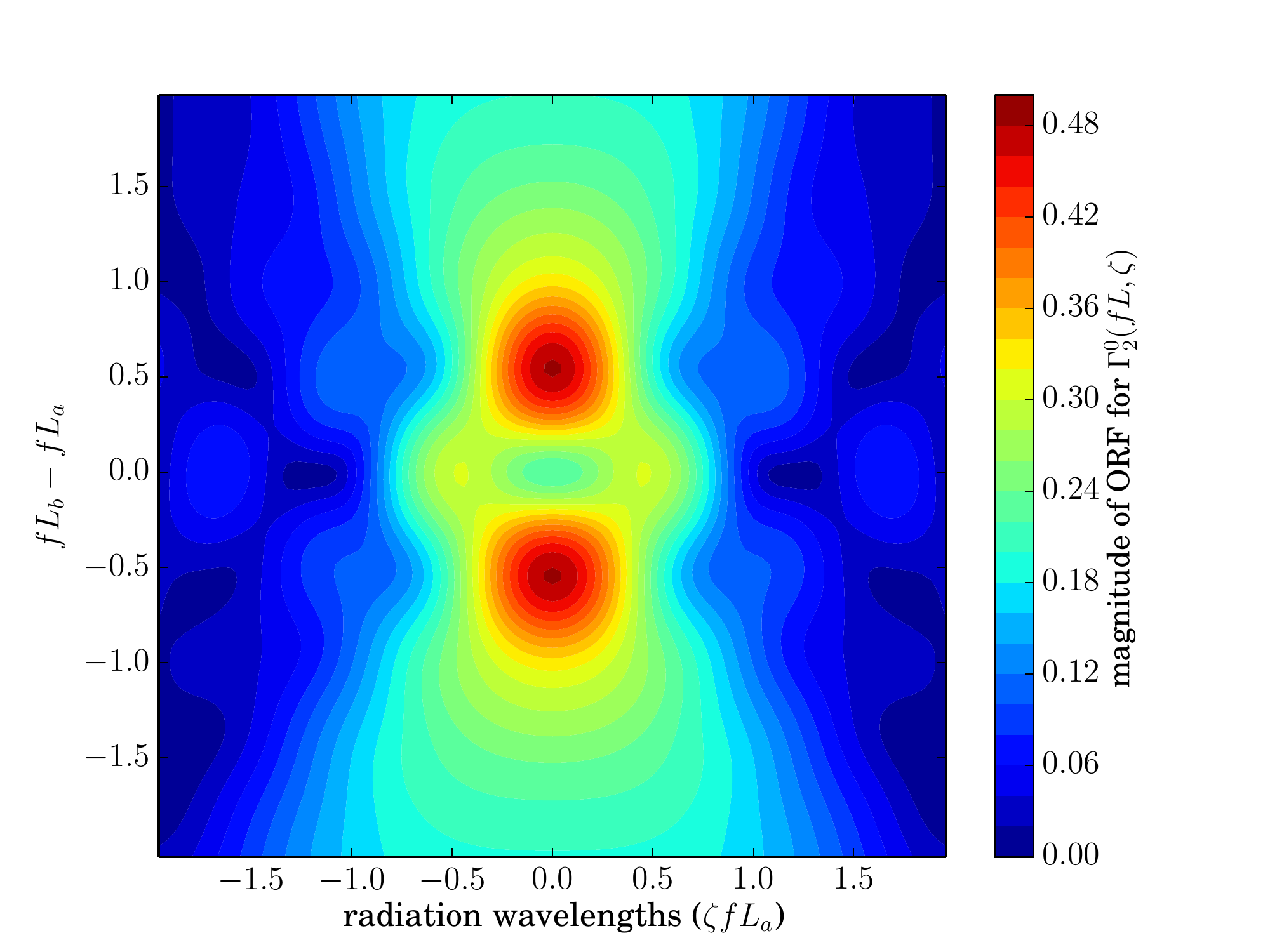}
        }\\ %  ------- End of the first row ----------------------%
                       \subfigure[%
	\, Magnitude of the quadrupole $^{(ab)}\Gamma^1_2(fL,\zeta)$ ORF.
        ]{%
	\label{fig:mag12}
           \includegraphics[width=0.48\textwidth]{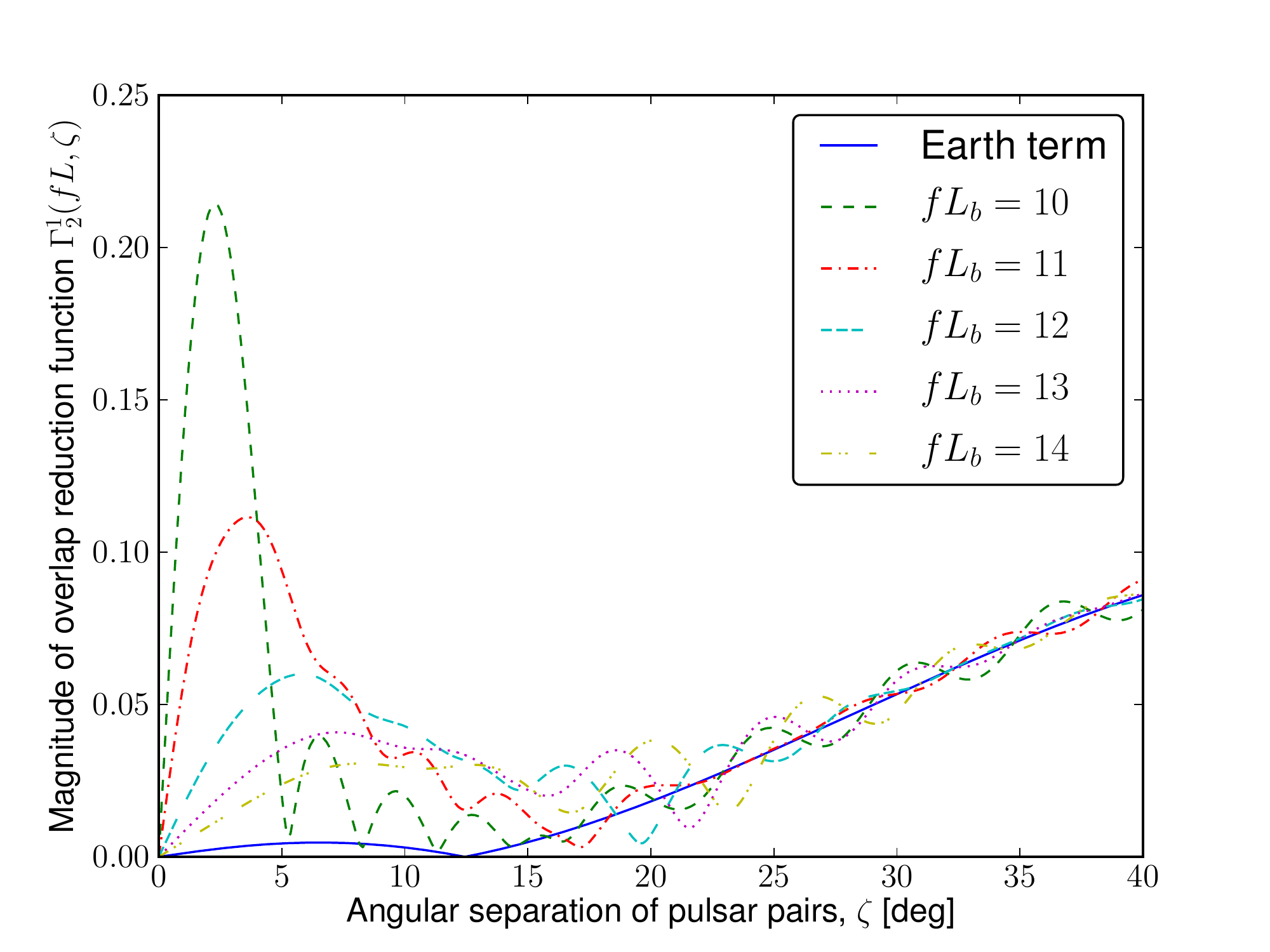}
        }
        \subfigure[%
        \,Strong pulsar term regime for $^{(ab)}\Gamma^1_2(fL,\zeta)$]{%
            \label{fig:contour12}
            \includegraphics[width=0.48\textwidth]{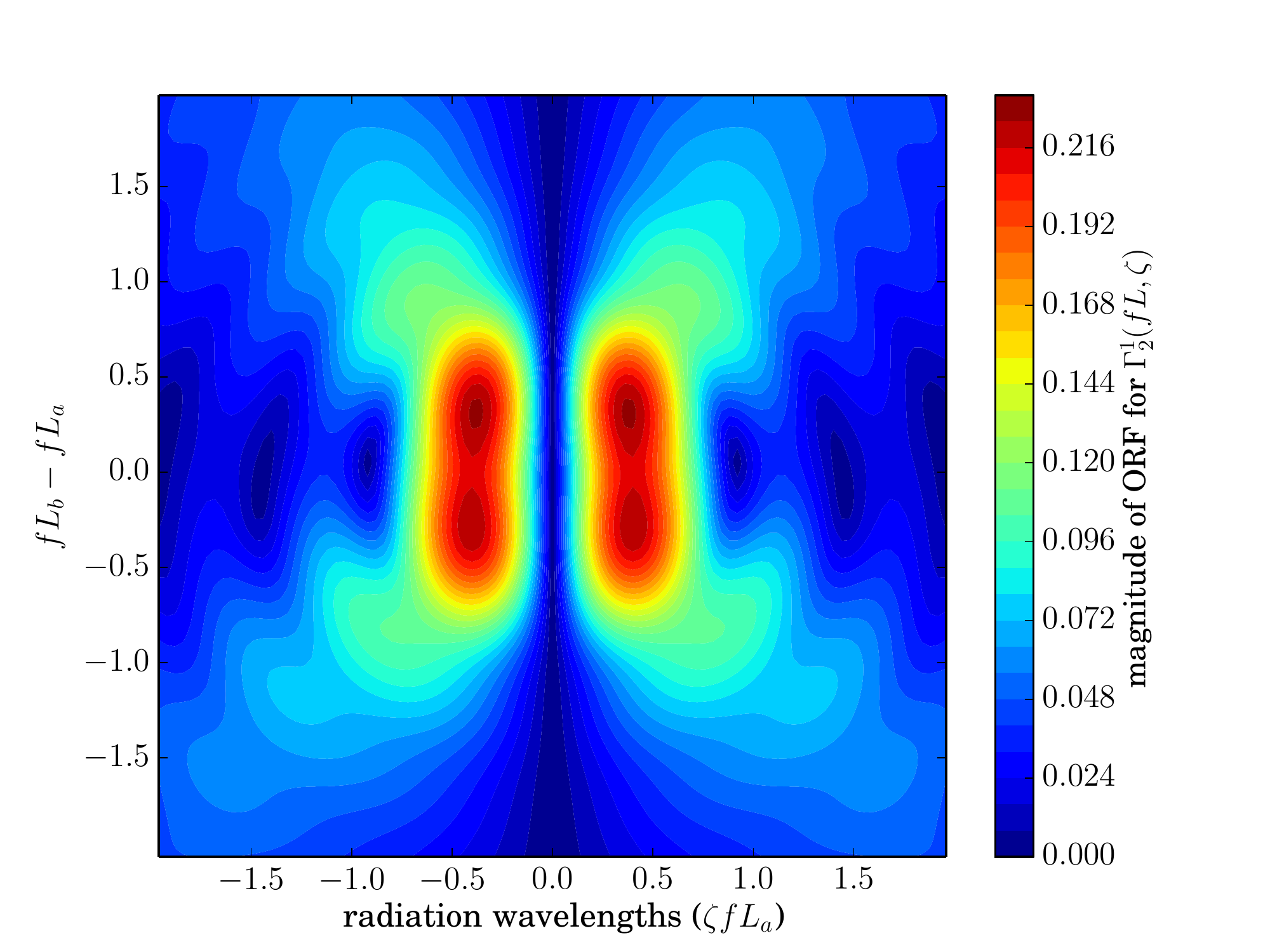}
        }
\\ %  ------- End of the first row ----------------------%
                       \subfigure[%
	\, Magnitude of the quadrupole $^{(ab)}\Gamma^2_2(fL,\zeta)$ ORF.
        ]{%
	\label{fig:mag22}
           \includegraphics[width=0.48\textwidth]{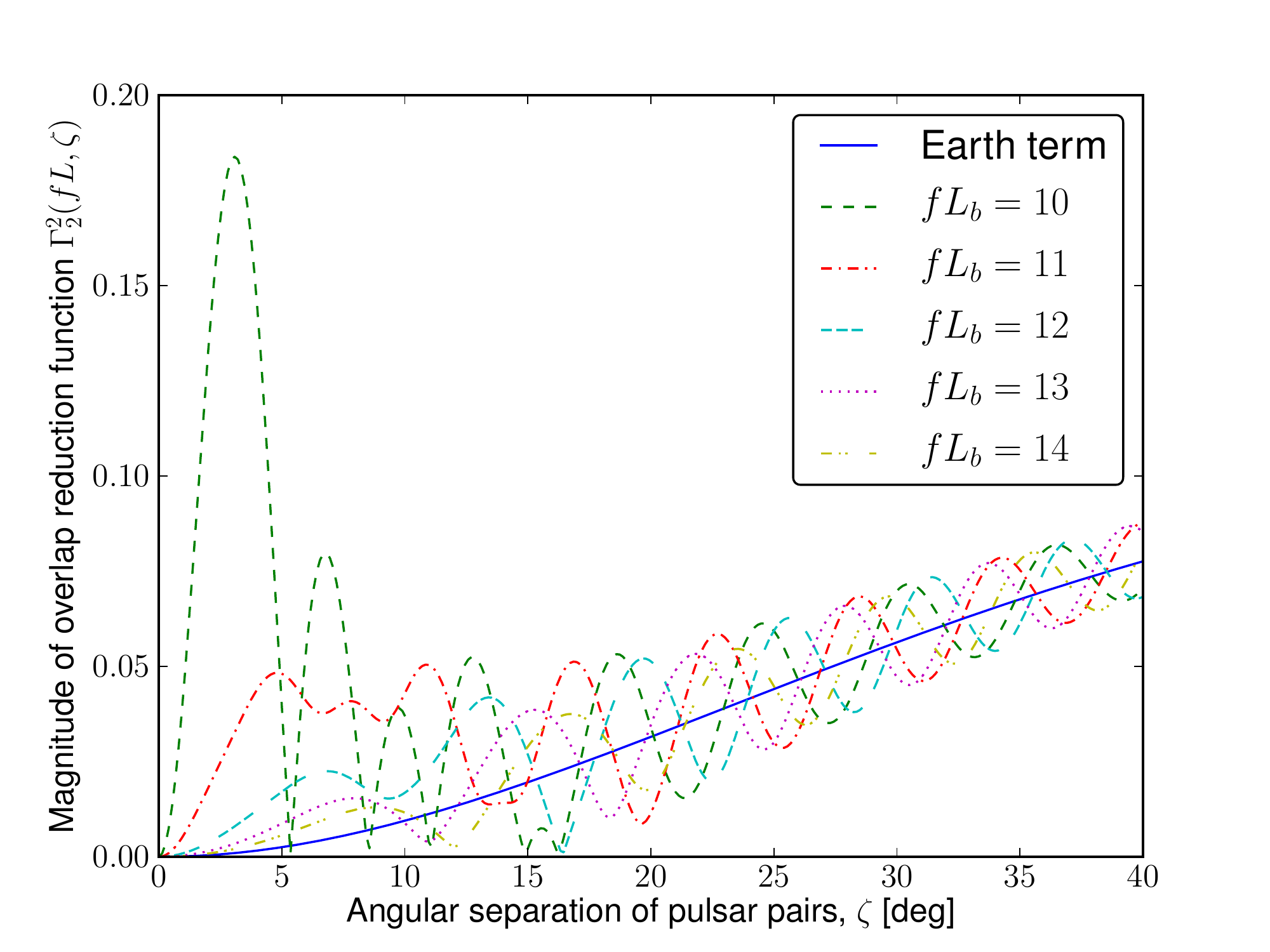}
        }
        \subfigure[%
        \,Strong pulsar term regime for  $^{(ab)}\Gamma^2_2(fL,\zeta)$]{%
            \label{fig:contour22}
            \includegraphics[width=0.48\textwidth]{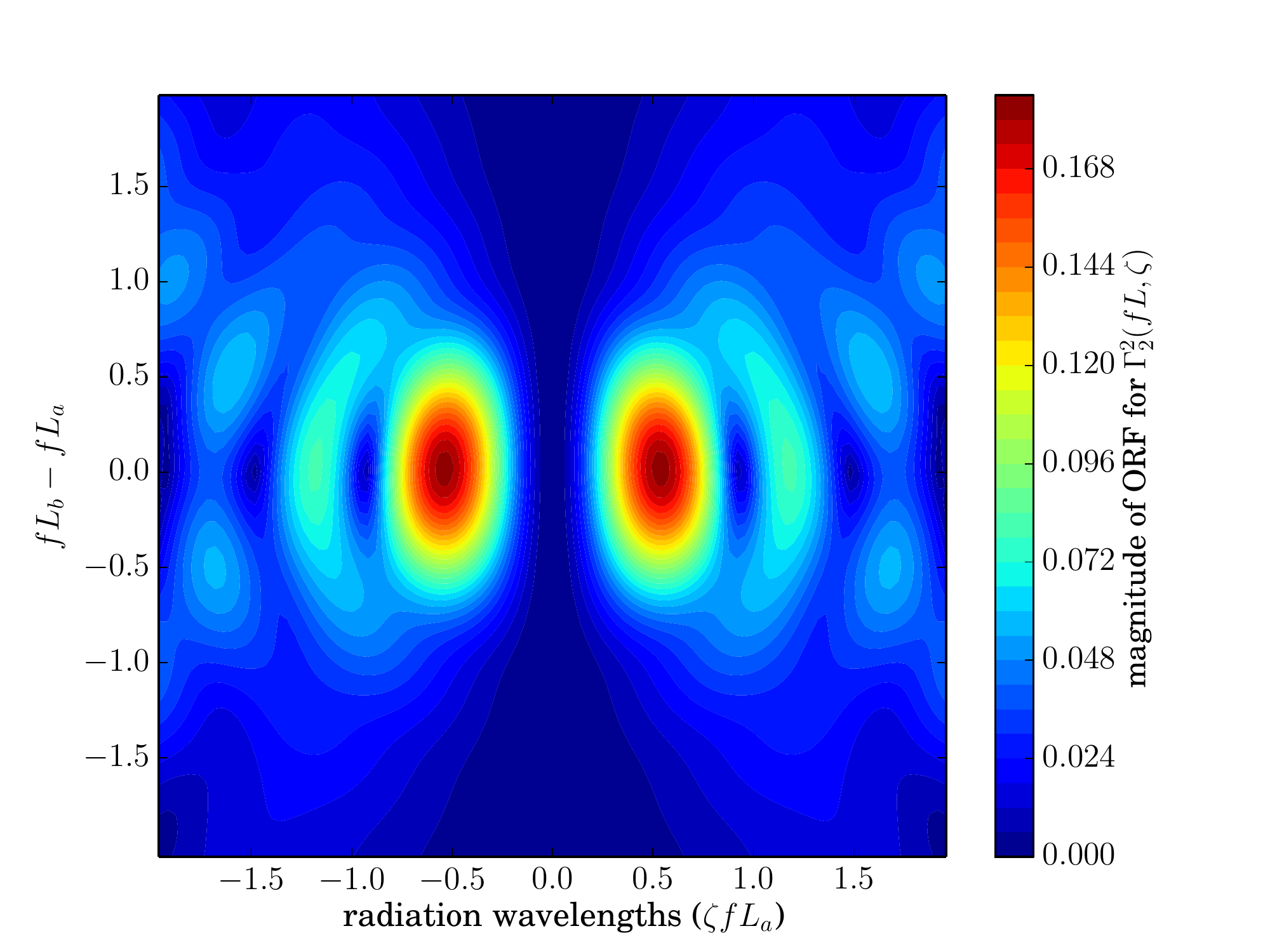}
        }

    \end{center}
    \caption{The effect of pulsar distance variations on the magnitude of the quadrupole overlap reduction functions, with $fL_a=10$ fixed. Panels on the left hand side are truncated at 40 degrees, as pulsar term oscillations rapidly converge to zero. Note that the maximum value of these ORFs is achieved for small, but non-zero, pulsar separations.} %
   \label{fig:fig22}
\end{figure*}

We probe the strong pulsar term regime-- where the pulsars are
separated by less than a few radiation wavelengths -- by continuously
moving pulsar $b$ towards or away from pulsar $a$ radially (along the $z$-axis). This change in distance is given by $fL_b-fL_a$, as shown in
Fig \ref{fig:pulsar_cartoon}. Pulsar $b$ is also moved
azimuthally away from $a$ by an angle $\zeta$, and
therefore the number of radiation wavelengths $b$ is from $a$ is
approximately given by $\zeta fL_a$. The effect of these continuous movements on the magnitude of the ORFs is shown in the contour plots in Figs \ref{fig:00-11}, \ref{fig:fig22}. The overall shape of the ORFs in the strong pulsar term regime is a function of the geometry of the pulsars and how they are aligned with the GW energy density, which is in turn described by the standard spherical harmonics $Y^m_l(\theta,\phi)$. A detailed explanation of the features seen in the contours in Figs \ref{fig:00-11} and \ref{fig:fig22} is given in Appendix~\ref{sec:features} and the most significant differences between the complete ORF, Eq. \eqref{eq:HDsol}, and the Earth-term-only ORF, Eq. \eqref{e:abGammalm1}, are highlighted in Table \ref{table1}. The ORFs are truncated at $\zeta\sim 40\Degs$, since the pulsar-term-induced oscillations at larger angular separations are smaller than the expected error of the average timing residual, which is roughly a few tens of percent~\cite{Jenet:2005, Anholm:2009, NANOGrav}.

We first study the magnitude of the isotropic ORF
$^{(ab)}\Gamma^0_0(fL,\zeta)$, which in the short wavelength
approximation is often
referred to as the Hellings and Downs
curve~\cite{HellingsDowns:1983}. Since this normalization is applied to
the isotropic ORF, it is also applied to the dipole and quadrupole
ORFs for consistency. 

The analysis continues with the study of the
dipole, $l=1$, $m=0,1$ and quadrupole $l=2$, $m=0,1,2$ ORFs. The $-m$
values of the ORFs are not explicitly shown, since in our reference
frame, described in Eq. \eqref{e:coord} and illustrated in Fig. \ref{fig:geometry},
\beq
^{(ab)}\Gamma^m_l(fL,\zeta)=(-1)^m {}^{(ab)}\Gamma^m_l(fL,\zeta)\, .
\eeq 
It may be surprising that all the ORFs are evaluated, since our previous studies~\cite{Mingarelli:2013} indicated that the $m=0$ ORFs were the most sensitive to the pulsar term. That study, however, only considered ORFs with non-zero auto-correlation values. In fact, small inter-pulsar distance variations will introduce correlated phase changes which are important for all the ORFs, as we show in Figs \ref{fig:00-11}, \ref{fig:fig22}.  A general formula for when to include the pulsar term is given at the end of this section.

%%%%%%
%
%%
% --------------  Include the figures for Gamma00 -------------------%
%%

%%
% --------------  Include the figures for l=2 -------------------%
%%

\subsection{The Hellings and Downs curve}
Firstly, we explore the behavior of the isotropic overlap reduction function when the pulsars are separated from each other by a few radiation wavelengths, either azimuthally or radially, cf. Fig \ref{fig:pulsar_cartoon}. 
The contour plot Fig \ref{fig:contour00} complements Fig \ref{fig:mag00} as it shows the continuous displacement of pulsar $b$ from pulsar $a$. We find that for a fixed pulsar $a$ with $fL_a=10$, the largest value of the ORF is achieved for $\zeta=0$ and $fL_b-fL_a=0$, as expected. Moreover, we find that the strongest pulsar term effects occur when pulsar $b$ located less than a radiation wavelength away from $a$, with the largest correlations occurring when pulsar $b$ is less than half a radiation wavelength from $a$. The magnitude of the pulsar-term-induced oscillations drops dramatically as pulsar $b$ is moved one radiation wavelength away from pulsar $a$. Moreover, the peak of these oscillations moves to the right as $fL_b$ increases, and the period of the oscillations increases. This behavior is present in all the ORFs, cf. Table \ref{table1} and Figs \ref{fig:00-11}, \ref{fig:fig22}. Indeed, it is clear that as $fL_b$ increases, the ORF converges to the Earth-term-only solution, the solid (blue) curve in Figs.~\ref{fig:00}, \ref{fig:mag00}. 

Our analysis of the isotropic ORF therefore indicates that the pulsar term only adds a significant additional piece to the standard, Earth-term-only ORF for equidistant pulsars within 10 radiation wavelengths from the SSB, separated by no more than half a GW wavelength, or equivalently $\zeta\lesssim3\Degs$ for $fL=10$, in agreement with Fig \ref{fig:mag00}. 
\subsection{The dipole overlap reduction function}
For $\Gamma^0_1(fL, \zeta)$, the largest contribution from the pulsar term arises from the scenario where $fL_a=fL_b=10$, seen in both Fig \ref{fig:mag01}, \ref{fig:contour01}. In Fig \ref{fig:mag01}, one can see that by moving pulsar $b$ one radiation wavelength to $fL_b=11$, the dashed-dot (red) curve, the additional contribution of the pulsar term is negligible. As $fL_b$ increases by one for each subsequent curve, it is clear that the pulsar term contribution converges to zero. Therefore the ORF becomes essentially an Earth-term only expression as the pulsars are separated by many radiation wavelengths (or large angles). 
In  Fig \ref{fig:contour01}, we find that the strong pulsar term region is extended in the $fL_b-fL_a$ direction, indicating that the pulsar term is important when pulsar $b$ is up to one radiation wavelength away (in the $z$ direction) from pulsar $a$. This strong pulsar term range is twice that of the isotropic ORF in the $z$-direction, but in terms of azimuthal radiation wavelengths, $\zeta fL_a$, the sensitivity is very similar to that of the isotropic ORF. The shape is due to a combination of geometric effects and the transverse nature of GWs, described in Appendix \ref{sec:features}.

For the $\Gamma^1_1(fL, \zeta)$ ORF, the largest contribution from the pulsar term arises from the scenario where $fL_a= fL_b$, as shown in Fig~\ref{fig:contour11}, but the maximum is achieved at a non-zero angular separation of $\zeta_\mathrm{max}=1.9\Degs$. Note that the fractional difference between the full ORF and the Earth-term-only ORF at $\zeta_\mathrm{max}$ is $49$! This ORF also differs from the ones studied so far in that the relatively large oscillatory behavior is present up to $\zeta \lesssim 20\Degs$. Moving pulsar $b$ one radiation wavelength to $fL_b=11$-- the dashed-dot (red) curve in Fig \ref{fig:mag11}-- the additional contribution of the pulsar term is still remarkable, with its peak at $\zeta_\mathrm{max}=3.5\Degs$, and a fractional difference between the full and Earth-term only ORF of $7$. 
The strong pulsar term region is extended in the $fL_b-fL_a$ direction, as it was for $\Gamma^0_1(fL, \zeta)$, with the exception of having no response at $\zeta=0$, see Fig \ref{fig:contour11}. The peak is centered on $\zeta fL_a\sim 0.5$ and extends to $\zeta fL_a\sim1$, which translates into important pulsar term features for $0\Degs< \zeta \lesssim 6\Degs$, in agreement with Fig \ref{fig:mag11}. The oscillations are slower to converge for this ORF, and therefore one may wish to include these additional correlated phase changes in stochastic GW background searches, up to $\zeta\sim 15 \Degs$ when $fL_b\sim10-12$, see Fig \ref{fig:mag11}.
\subsection{The quadrupole overlap reduction function}
Here we examine how varying the distances to pulsars in a PTA affects
the behavior of the $l=2$, $m=0,1,2$ quadrupole ORFs
$^{(ab)}\Gamma^m_2(fL,\zeta)$. The key figure for this analysis is Fig \ref{fig:fig22}. As before, we fix $fL_a=10$ and vary
$fL_b$ from 10 to 14. The values of $fL_b$ were calculated up to
$fL_b=20$, however as before, these additional curves converged to
zero very quickly, providing little insight. 

Starting with $^{(ab)}\Gamma^0_2(fL,\zeta)$, Fig \ref{fig:mag02}, the two main curves of interest are $fL_b=10$ and $fL_b=11$. 
Although the ORF is twice the Earth-term for $fL_a=fL_b=10$ at $\zeta=0$, as expected \cite{Mingarelli:2013}, the maximum value of the magnitude of this ORF is at $\zeta_\mathrm{max}=2.4\Degs$, where it is triple the value of the Earth term, with a fractional difference of $\sim 2$. Moving pulsar $b$ a radiation wavelength away from pulsar $a$, corresponding to $fL_b=11$ in Fig \ref{fig:mag02}, the auto-correlation is three times larger than the Earth-term expression. Moreover, for $fL_b=11$, the magnitude of the auto-correlation of $^{(ab)}\Gamma^0_2(fL,\zeta)$ is {\it larger} than the $fL_a=fL_b=10$ case. 

The full $m=1$ and $m=2$ quadrupole ORFs also feature a remarkable departure from the Earth-term-only expression for pulsars separated by less than a radiation wavelength, and converge more slowly to the Earth-term-only ORF (solid blue curve), Figs \ref{fig:mag12}, \ref{fig:contour12}, \ref{fig:mag22}, \ref{fig:contour22}. For $^{(ab)}\Gamma^1_2(fL,\zeta)$, the pulsar term contribution is important for pulsars separated by up to $6\Degs$ and for $^{(ab)}\Gamma^2_2(fL,\zeta)$, with $fL_a=fL_b=10$, the fractional difference between the Earth-term-only and full ORF is 188 at $\zeta_\mathrm{max}=3.1\Degs$.

\subsection{Summary}
From Table \ref{table1}, one can note that the largest values for the magnitude of the ORFs is achieved when equidistant pulses are separated by small angles. However, pulsars separated by up to two radiation wavelengths (denoted below as $\lambda=2$) could contribute additional correlated phase terms to the ORF which may need to be modelled, depending on the magnitude of the error bars for each point on the curve, which in turn depend on the observations. The correlated phase changes are therefore important for pulsars separated by
\beq
|L_a -L_b|=19\, \left(\frac{\lambda}{2} \right) \left( \frac{f}{10^{-9}~\mathrm{Hz}}\right)^{-1}\, \mathrm{pc} \, .
\eeq

One can also use the Law of Cosines to estimate when the pulsar term should be included. As above, let $\lambda$ be the number of radiation wavelengths separating a pulsar pair. Then by the Law of Cosines,

\beq
\label{eq:LawOfCos}
\lambda^2<(fL_a)^2+(fL_b)^2-2(fL_a)(fL_b)\cos\zeta \, ,
\eeq
or
\beq
\label{eq:Cos}
\cos\zeta < \frac{1}{2(fL_a) (fL_b)}\left[  \left(fL_a\right)^2 + \left(fL_b\right)^2-\lambda^2\right]\, .
\eeq
For an isotropic stochastic GW background, we found that the pulsar term should be considered for pulsars separated by half a radiation wavelength or less, see Figs \ref{fig:mag00}, \ref{fig:contour00} and Table \ref{table1}.  Therefore by Eqs. \eqref{eq:LawOfCos} and \eqref{eq:Cos} with $fL_a=fL_b=10$ and $\lambda=1/2$, the pulsar term should be considered for angular separations less than

\beq
\label{eq:IsoEst}
\zeta \leq \arccos(799/800)~\mathrm{rads}\sim 2.86 \Degs.
\eeq
Note that the non-linearity of Eq. \eqref{eq:Cos} prevents us from writing down a straightforward scaling relation as a function of $fL_a$, $fL_b$ and $\lambda$.

\begin{table*}
\begin{center}
\begin{tabular}{|cccccc|cccccc|}
\hline
ORF												& $fL_b$ 			&	  $~\zeta_{\mathrm{max}}$	& Full ORF	& 	ET ORF &	Frac Diff 		&	ORF		& $fL_b$ 			&	  $~\zeta_{\mathrm{max}}$	& Full ORF	& 	ET ORF &	Frac Diff \\
\hline
\multirow{3}{*}{$\Gamma^0_0$} 	& 		$10$ 		& 	$~0.0\Degs$				&			$1.0$ 	&		$5.0\times10^{-1}$		&	$~~1$ 		& \multirow{3}{*}{$\Gamma^0_2$} 	& 		$10$ 		& 	$~2.4\Degs$				&			$3.0\times10^{-1}$ 	&		$1.0\times10^{-1}$		&	$~~2$ \\	
													&		$11$			& 	$~0.8\Degs$				&			$5.7\times 10^{-1}$	&		$4.8\times10^{-1}$		&	$~~0.2$ 		&	&	$11$			& 	$~0.0\Degs$				&			$3.3\times 10^{-1}$	&		$1.1\times10^{-1}$		&	$~~2$\\		
													
													&		$12$			&	$~1.1\Degs$				&			$5.2\times10^{-1}$	&		$4.8\times10^{-1}$		& 	$~~0.1$		&	&	$12$			&	$~2.4\Degs$				&			$1.9\times10^{-1}$	&		$1.0\times10^{-1}$		& 	$~~0.9$	\\	
\hline
\multirow{3}{*}{$\Gamma^0_1$} 	& 		$10$ 		& 	$0.0\Degs$				&			$8.6\times10^{-1}$ 	&		$4.3\times10^{-3}$		&	$~~1$ &\multirow{3}{*}{$\Gamma^1_2$} 	& 		$10$ 		& 	$~2.3\Degs$				&			$2.1\times10^{-1}$ 	&		$2.5\times10^{-3}$		&	$~84$	\\
													&		$11$			& 	$0.5\Degs$				&			$5.5\times 10^{-1}$	&		$4.2\times10^{-3}$		&	$~~0.3$		&	&	$11$			& 	$~3.6\Degs$				&			$1.1\times 10^{-2}$	&		$3.6\times10^{-3}$		&	$~30$ 	\\	
													&		$12$			&	$~1.1\Degs$				&			$4.8\times10^{-1}$	&		$4.1\times10^{-1}$		 	&$~~0.2$	&	&	$12$			&	$~5.6\Degs$				&			$6.0\times10^{-2}$	&		$4.6\times10^{-3}$		& 	$~12$	\\	
\hline
\multirow{3}{*}{$\Gamma^1_1$} 	& 		$10$ 		& 	$~1.9\Degs$				&			$2.5\times10^{-1}$ 	&		$4.9\times10^{-3}$		&	$49$ & \multirow{3}{*}{$\Gamma^2_2$} 	& 		$10$ 		& 	$~3.1\Degs$				&			$1.8\times10^{-1}$ 	&		$9.7\times10^{-4}$		&	$188$	\\
													&		$11$			& 	$~3.5\Degs$				&			$6.9\times 10^{-2}$	&		$9.0\times10^{-3}$		&	$~7$&		&	$11$			& 	$~4.7\Degs$				&			$4.8\times 10^{-2}$	&		$2.2\times10^{-3}$		&	$~21$ 	\\	
													&		$12$			&	$5.5\Degs$				&			$4.2\times10^{-2}$	&		$1.4\times10^{-2}$		& 	$~2$&		&$12$			&	$13.2\Degs$				&			$4.2\times10^{-2}$	&		$1.6\times10^{-2}$		& 	$~~2$	\\	
\hline
\end{tabular}
\end{center}

\caption {Here we list the largest fractional difference of the magnitude of the full ORFs (F) and Earth-term-only ORFs (ET), see Figs \ref{fig:00-11}, \ref{fig:fig22}, and report the angle at which this maximum value was achieved, $\zeta_\mathrm{max}$.  The value of $fL_a$ is fixed at 10. The fractional difference of the magnitude of the ORF is $|\mathrm{F}-\mathrm{ET}|/\mathrm{ET}$, rounded to the nearest integer unless it is less than one. Note that for $\zeta_\mathrm{max}=0$ in $\Gamma^0_{l=0,1}$ the pulsar term is adequately modelled by $(1+\delta_{ab})$, see Eq. \eqref{e:abGammalm1}. }
\label{table1}
\end{table*}

In the following section, we give an approximation to $\kappa_{ab}(fL,\zeta)$ which can be used for small angular pulsar separations in order to model this effect.

\section{Small angle approximation of the pulsar terms}
\label{s:pulsarterm}

In Sec \ref{s:distance} we showed that the pulsar term is important to
include in the evaluation of all the ORFs if the pulsars are separated
by less than a few radiation wavelengths, see Table \ref{table1} for
details. Motivated by the possibility of having pulsars separated by
such a small angle in future PTA experiments, we give a small angle approximation
of the pulsar term, up to  $\cal{O}$$(\zeta^2)$ which
closely follows the true behavior of the complete isotropic
ORF. This approximation can be easily integrated into stochastic GW
background search pipelines, and will be faster to evaluate than the full expression. 
 
\begin{figure}
	\centering
	\includegraphics[scale=.45]{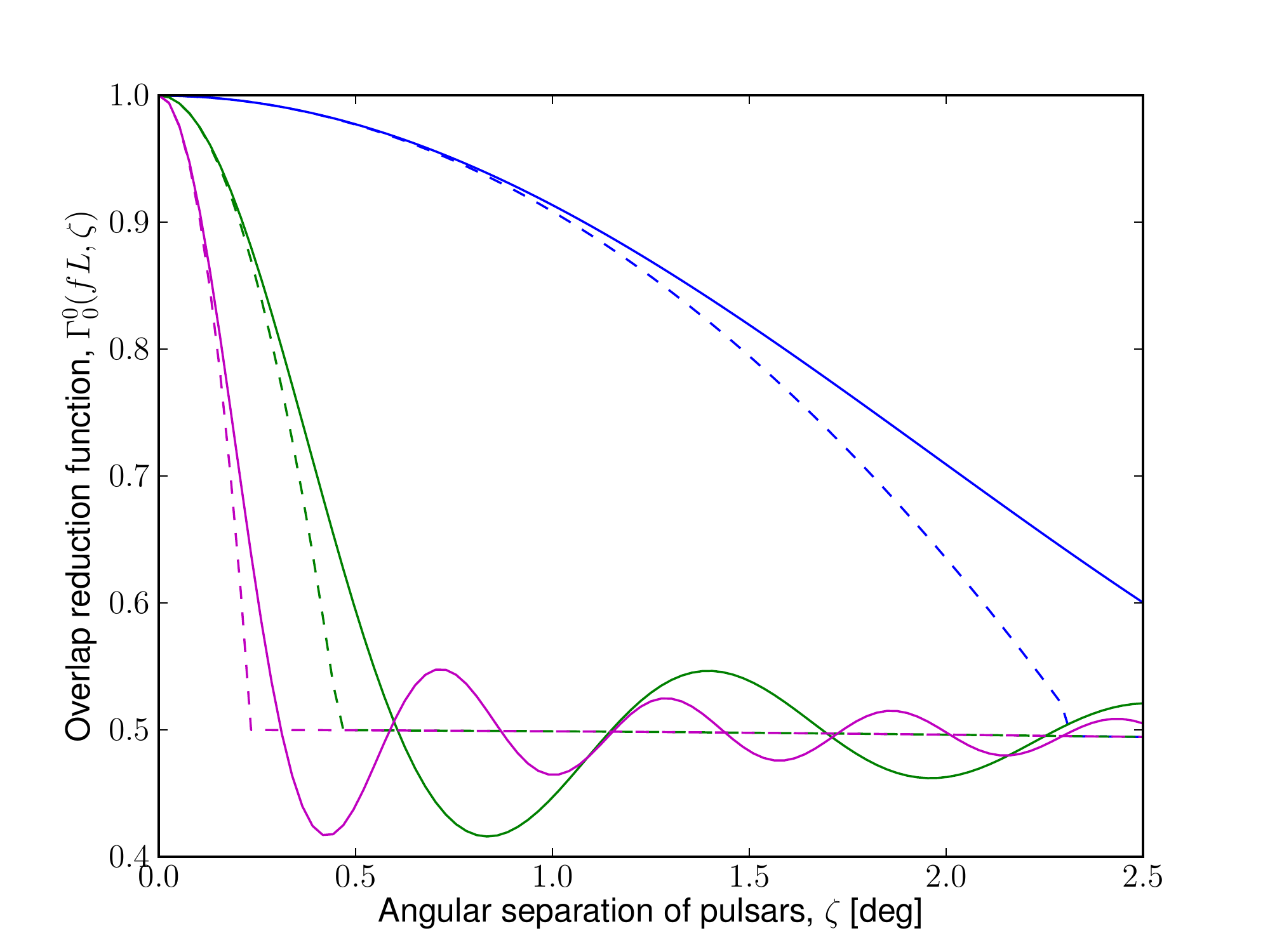}
	\caption{The small angle approximation of the isotropic ORF compared to the full ORF. Moving from right to left: the solid curve is the full ORF and the dashed one is the approximation given by Eq. \eqref{eq:RfullKappa} for $fL=10$ (online blue), $fL=51.2$ (green), and $fL=100$ (magenta). The approximation appears to hold for $\zeta \leq 2.3\Degs$ for $fL\geq 10$, Eq. \eqref{eq:approxZeta}. Afterward the ORF reverts to the Earth-term-only solution, which appears flat due to the small range of angles considered.
	} 
	\label{fig:approx}
\end{figure}

Since the pulsar term, Eq. \eqref{eq:kappa}, is not a function of
angular distribution of the GW energy density, this approximation can
be used for all PTA ORFs, however it is advised to extend the
approximation to $\cal{O}$$(\zeta^3)$ for $l\ge 1$. We show how this
approximation compares to the full isotropic ORF for $fL_{a=b}=10,
100$, and 51.2 as a non-integer example, see Figure \ref{fig:approx}. 

 \begin{widetext} 
Working in the computational frame, as defined in Eq.~\eqref{e:coord} and shown in Fig \ref{fig:geometry}, we write down the full expression for the isotropic ORF,  Eq.~(\ref{eq:HDsol}), with $Y^0_0(\theta,\phi)=1/\sqrt{4\pi}$:

\beq
	{}^{(ab)}\Gamma^0_0(\zeta)=\frac{1}{\sqrt{4\pi}}\int_{S^2}d\hat\Omega\,  \left[ 1 - e^{i 2 \pi f L_a (1+ \hat \Omega \cdot \hat p_a)} \right] \left[ 1 - e^{-i 2 \pi f L_b (1+ \hat \Omega \cdot \hat p_b)}\right] F^+_a(\hat\Omega)F^+_b(\hat\Omega).
        \label{eq:reductionSpheric}
\eeq

Recall that for anisotropic ORFs, one will need to rotate the pulsars back into the cosmic rest frame from the``computational frame'' using Wigner D matrices~\cite{Mingarelli:2013, TaylorGair:anisotropy, Gair:2014}.  This is because for anisotropic stochastic GW backgrounds, the position of the pulsar pair with respect to the background GW energy density matters.

In order to simply the following notation, we define
\bea
\label{eq:MandN}
M&=&2\pi fL_a(1+\cos\theta),\\ 
N&=&2\pi  fL_b(1+\cos\theta\cos\zeta+\sin\theta\sin\zeta\cos\phi) \, ,
\eea
and write $\kappa_{ab}(f,\hat\Omega)$ in terms of trigonometric functions, separating real and imaginary parts:
\bea
\label{eq:kappaExpand} 
\kappa_{ab}(f,\hat\Omega)&=&(1-e^{iM})(1-e^{-iN})\, ,\nonumber\\
&=&\cos(M-N)-\cos M-\cos N+1+i[\sin(M-N)-\sin M+\sin N].
\eea

For small angles, one can approximate $\kappa_{ab}(f,\hat \Omega)$ with a Taylor Series about $\zeta=0$. Since the following expression is not a straightforward function of 
either $M$ or $N$, we do not use these definitions, although they will be re-introduced when convenient.

To order $\mathcal{O}(\zeta^2)$, the real part of this expansion is:
\bea \label{eq:ReExpansion}
\Re(\kappa_{ab}(f,\hat\Omega))& \approx &
%leading order
1-\cos[2\pi fL_a(1+\cos\theta)]-\cos[2\pi fL_b(1+\cos\theta)]+\cos[4\pi (fL_a-fL_b)\cos^2(\theta/2)]\nonumber\\
% order zeta
&+& \zeta \,4 \pi fL_b \sin \theta \cos \phi  \cos \left[2 \pi  (fL_a-2 fL_b) \cos ^2(\theta/2)\right] \sin [\pi fL_a (\cos\theta +1) \nonumber \\
% order zeta^2
&-&\zeta^2 \,\pi fL_b \left\{2 \pi fL_b \sin ^2\theta  \cos ^2\phi  \left[\cos \left(4 \pi  [fL_a-fL_b] \cos ^2(\theta/2)\right)-\cos (2 \pi  fL_b [\cos \theta +1])\right]  \right. \nonumber\\
 &\quad& + \left .\cos \theta \left(\sin \left[4 \pi (fL_a-fL_b) \cos ^2(\theta/2)\right]+\sin [2 \pi  fL_b(\cos \theta +1)]\right)\right\} +\cdots
\eea
and the imaginary part is
\bea
\label{eq:ImExpansion}
\Im(\kappa_{ab}(f,\hat\Omega)) &\approx& 
%leading order
\sin \left[4 \pi  (fL_a-fL_b) \cos ^2(\theta/2)\right]-\sin [2 \pi fL_a (\cos\theta+1)]+\sin [2 \pi fL_b (\cos\theta+1)] \nonumber \\
% order zeta
&+& \zeta 4 \pi  fL_b \sin \theta  \cos \phi  \sin [\pi fL_a (\cos\theta +1)] \sin \left[2 \pi  (fL_a-2 fL_b) \cos ^2(\theta/2)\right]\nonumber \\
% order zeta^2
&-& \zeta ^2 \pi fL_b \left\{2 \pi  fL_b \sin ^2\theta  \cos ^2\phi  \left(\sin \left[4 \pi  (fL_a-fL_b) \cos ^2(\theta/2)\right]+\sin [2 \pi  fL_b (\cos \theta +1)]\right)\right. \nonumber\\
 &\quad& + \left .\cos \theta  \left(\cos [2 \pi  fL_b (\cos \theta +1)]-\cos \left[4 \pi  (fL_a-fL_b) \cos ^2(\theta/2)\right]\right)\right\} +\cdots \, .
\eea

Many of the ORFs share the feature that the pulsar term is most notable when when the pulsars are equidistant from the SSB. In this case, $L_a=L_b=L$, and Eqs.~\eqref{eq:ReExpansion}, \eqref{eq:ImExpansion} are significantly simplified: 
\bea
\label{eq:RfullKappa}
\kappa_{ab}(f,\hat\Omega)&\approx& 2 - 2 \cos M+2\zeta\pi fL\sin M \cos\phi\sin\theta+\zeta^2\pi fL[-\cos\theta\sin M + 2\pi fL(\cos M-1)\sin^2\theta\cos^2\phi]\nonumber\\
&+& i\zeta2\pi fL\sin\theta\cos\phi(\cos M-1)+i\zeta^2\pi fL[\cos\theta(1\!-\!\cos M)\!-\!2\pi fL\sin M\!\sin^2\theta\cos^2\!\phi ]+\cdots \, .
\eea
\end{widetext}
When $\zeta=0$, Eq. \eqref{eq:RfullKappa} simplifies to $2-2\cos M$. At a glance, one may assume it is safe to ignore the $-2\cos M$ term since it is suppressed by a factor of at least $1/fL$ when integrated in the evaluation of the ORF.  In fact  in Appendix C of~\cite{Gair:2014}, it is shown that this additional $2\cos M \propto1/(fL)^2$ for the isotropic ORF, and so may safely be ignored for $fL\geq 10$. Section \ref{s:closepulsar} gives more details on this. 

As one may expect, the imaginary part of $\kappa_{ab}(f,\hat\Omega)$ vanishes for the $L_a=L_b$ isotropic case but is otherwise non-vanishing. This fact is somewhat masked by the use of the magnitude of the ORFs, instead of the individual real and imaginary components.

Since the stochastic GW background is likely to be highly isotropic at low frequencies, we verify this approximation numerically against the full isotropic ORF. We find that Eq.~\eqref{eq:RfullKappa} is a good approximation for the isotropic ORF when 
\beq
\label{eq:approxZeta}
\zeta\leq 2.3\Degs \left(\frac{10}{fL}\right), 
\eeq
as seen in Fig.~\ref{fig:approx}. Note that these values fall in the strong pulsar term regime predicted by our estimates in Eq. \eqref{eq:IsoEst}.

Using the pulsars found in the IPTA Mock Data Challenge 1 \cite{IPTAmdc}, we found that the smallest separation between pulsar pairs was $\zeta\sim3.5\degree$ for pulsars J1853+1303 and J1857+0943. Although this angle is indeed small, the distances to these pulsars found in the ATNF catalogue \cite{ATNF} are 1.6~kpc and 0.9~kpc, respectively, meaning that their $fL$ values in the low frequency limit are $168$ and $90$, respectively. Therefore the Earth-term only ORF is still a reasonable approximation for pulsar pairs in the IPTA mock data challenge.

\begin{figure*}
\begin{center}
        \subfigure[\,Real pulsar term contribution, $\Re$(Full - Earth-term-only ORF) for $^{(ab)}\Gamma^m_l(fL,\zeta)$ where $m=0$, $l=0,1,2$] {%
		\label{fig:realf1}
		\includegraphics[scale=0.42]{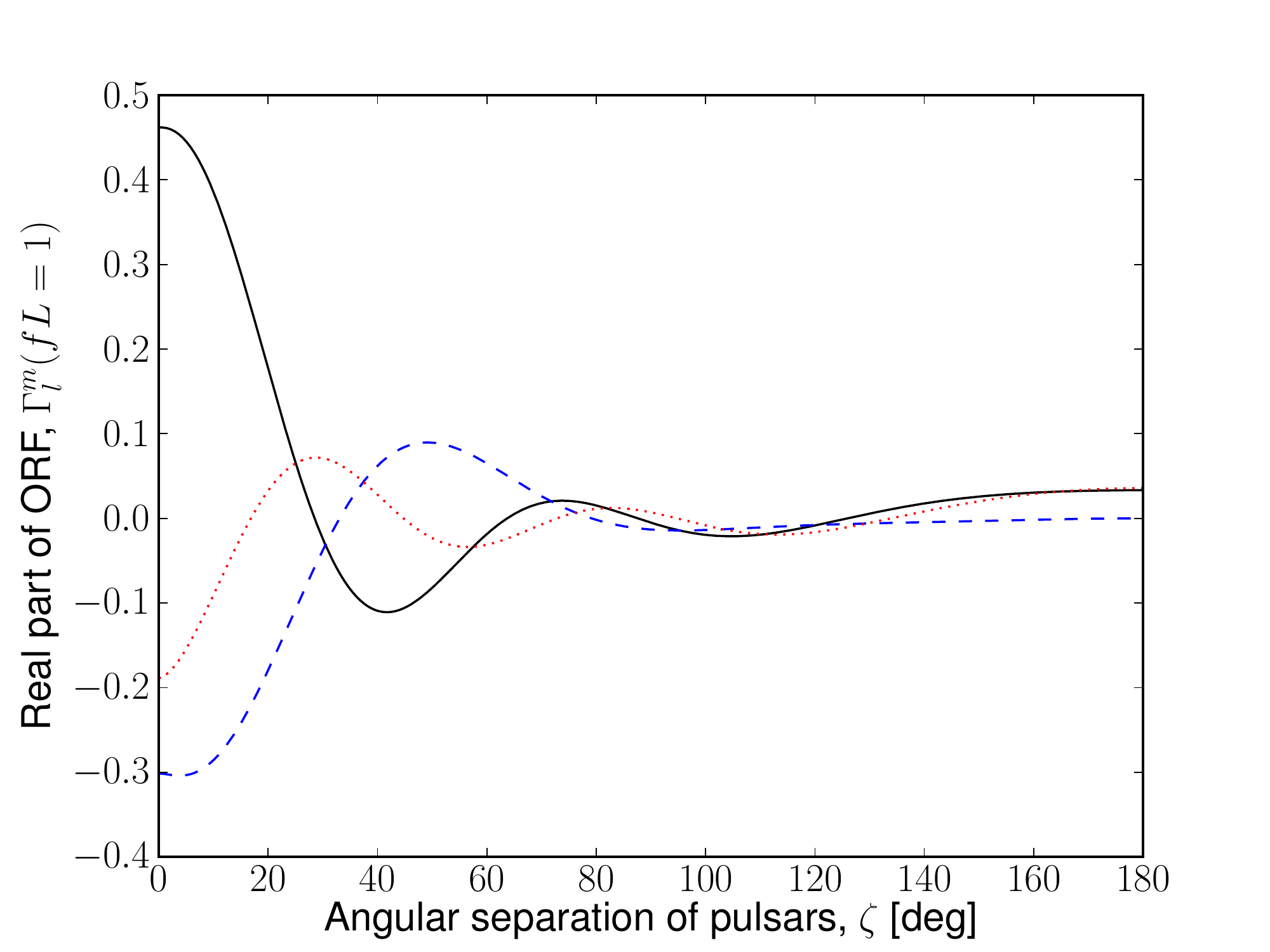}
        }%
	\subfigure[\, Imaginary part of  $^{(ab)}\Gamma^m_l(fL,\zeta)$ where $m=0$, $l=0,1,2$]{
	\label{fig:imagFL1}	
	\includegraphics[scale=0.42]{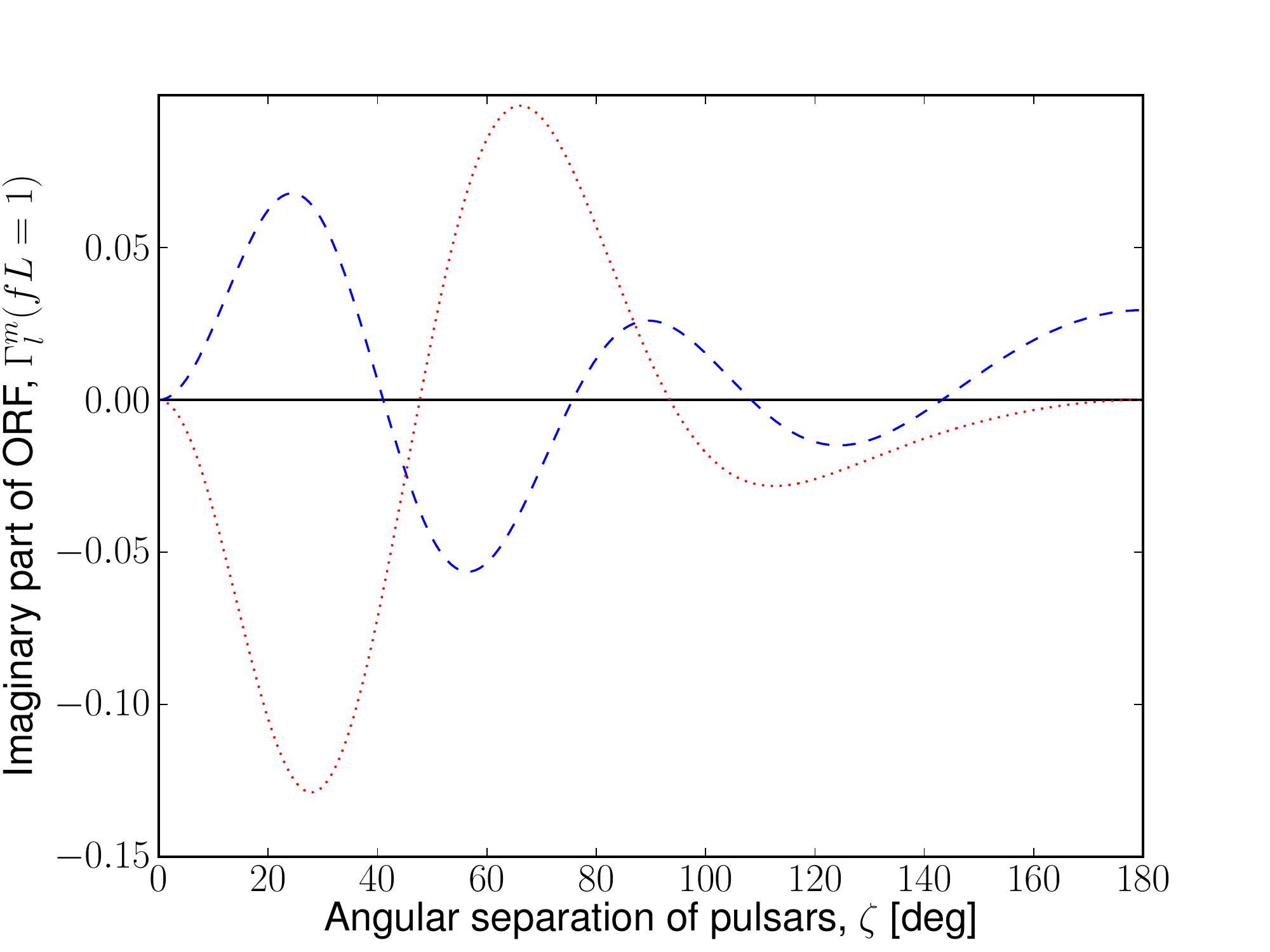}
	}
	\subfigure[\, Magnitude of the overlap reduction functions]{
	
	\label{fig:magFL1}
	
	\includegraphics[scale=0.42]{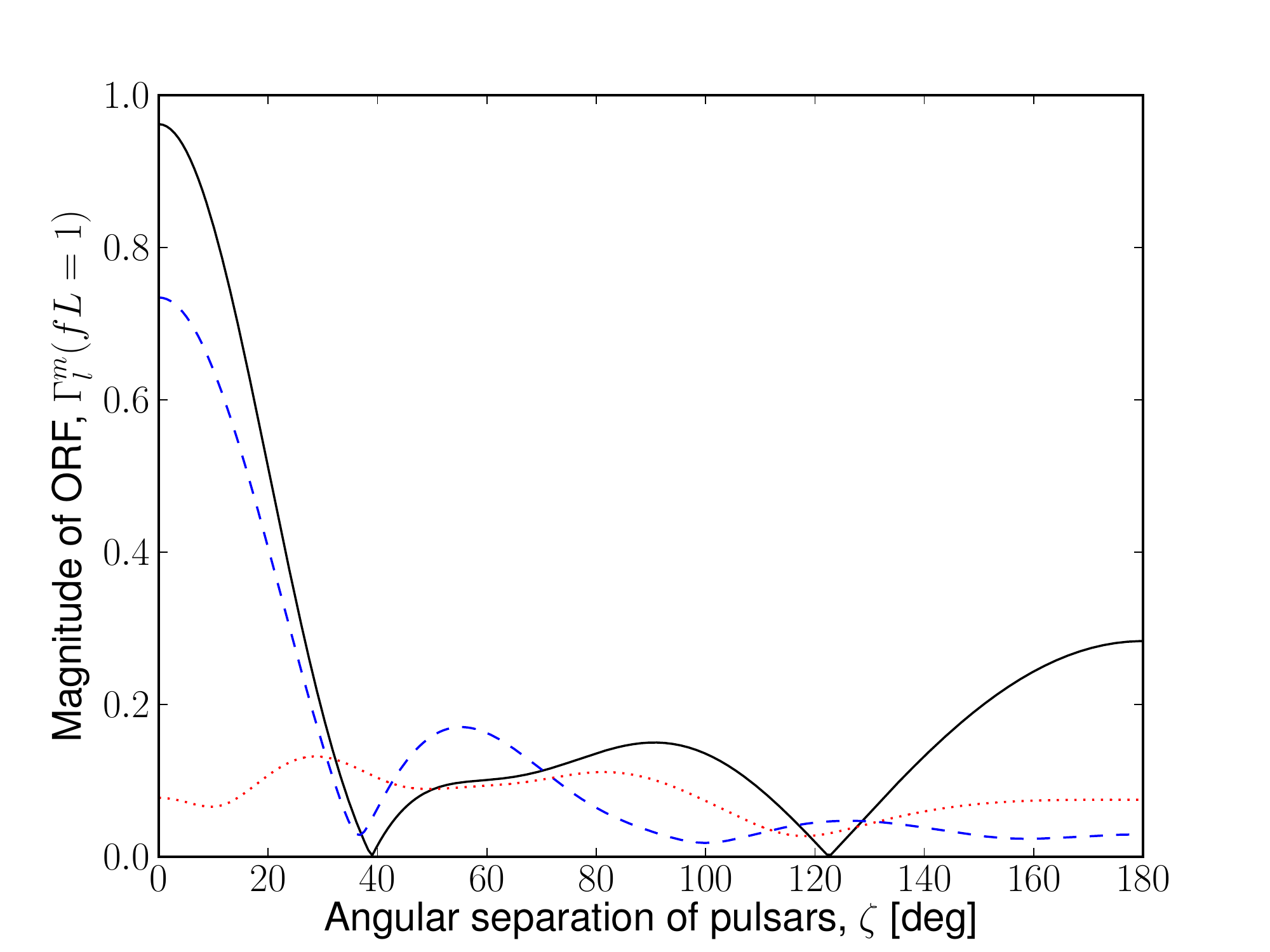}
	}
	
\end{center}
	\caption{In all panels $^{(ab)}\Gamma^0_0(fL,\zeta)$ is the solid curve, $^{(ab)}\Gamma^0_1(fL,\zeta)$ is the dashed curve (blue) and $^{(ab)}\Gamma^0_2(fL,\zeta)$ is the dotted curve (red). (a) The behavior of the pulsar term only when $L_a=L_b=L$ and $fL=1$ for the aforementioned ORFs.  This is found by subtracting the Earth-term solution from the numerically integrated ORF.  
(b) The imaginary part only of $^{(ab)}\Gamma^0_0(fL,\zeta)$, $^{(ab)}\Gamma^0_1(fL,\zeta)$ and $^{(ab)}\Gamma^0_2(fL,\zeta)$ when $fL=1$. As there is no imaginary part in the computational frame where the Earth-term is calculated, we cannot display the difference as is done in panel (a). Note that these imaginary values are only a factor a few smaller than their real counterparts, with the exception of $^{(ab)}\Gamma^0_0(fL,\zeta)$, where the imaginary part is zero. Moreover, they do not quickly converge to zero as in previous cases for $fL\geq10$.}
\label{fig:fLisOne}
\end{figure*}

\section{Correlated phase changes for pulsars within a radiation wavelength of the SSB}
\label{s:closepulsar}
At the time of writing, the ATNF pulsar catalogue \cite{ATNF} lists 16 pulsars which are closer than 300~pc, and three which are only 160 pc away. 
Statistically, neutron stars could be as close as 40~pc~\cite{Lorimer:2001}, with the closest known neutron star RX J185635-3754 at a distance of $61\pm9$~pc, discovered in the ROSAT all-sky survey \cite{WWN:1996}.
Our results suggest that the pulsar term can only be ignored if the distance between the pulsars is larger than a radiation wavelength (depending on the ORF), and/or $\zeta fL\geq 1$, cf. Figs \ref{fig:00-11}, \ref{fig:fig22} and Table \ref{table1}, with the exception of the auto-correlation which must always be considered. Current astrophysical constraints place a lower limit of $fL=10$, this limit may decrease as more nearby pulsars are found and added to PTAs in the future.  It may still be unlikely to discover pulsars which are closer than 100 pc, however for completeness we investigate the behavior of the ORFs when $fL\sim1$, i.e. when the SSB and the pulsar are separated by only one radiation wavelength. The results are shown in Fig \ref{fig:fLisOne}.

From the results shown in Fig \ref{fig:fLisOne}, it is clear that if $fL\sim 1$ the pulsar term has a very large effect on the ORF for pulsar pairs separated by $\zeta\leq90\Degs$. There are a few interesting features in these curves: firstly we note that $-2\cos M$ term from Eq. \eqref{eq:RfullKappa} affects the auto-correlation of all three ORFs. This can be seen by comparing the $\zeta=0$ points in Figs \ref{fig:mag00}, \ref{fig:mag01}, \ref{fig:mag02} to Fig \ref{fig:realf1} or equivalently Fig \ref{fig:magFL1}, since there is no imaginary part for the auto-correlation, see Fig \ref{fig:imagFL1}. Fig \ref{fig:magFL1} clearly shows that auto-correlation of the isotropic ORF is less than one. It is also interesting to note that the magnitude of $^{(ab)}\Gamma^0_2(fL,\zeta)$ shows little variation across the range of $\zeta$. Since the investigation of $fL\sim 1$ ORFs is purely speculative at the moment,  further detailed analyses are left for future investigations.

Tying together the small angle approximation with the small $fL$ considerations, we now give an example based on current pulsars in the ATNF catalogue. J0030+0451 is currently an IPTA pulsar and is 0.24~kpc away and its relatively close neighbor, J010-1431, is 0.13~kpc away. The latter is not currently being timed for PTA purposes, but is used here as an illustrative example. Assuming the low frequency limit for a PTA, we set $f=10^{-9}$~Hz, we find that $fL_a=24$ and $fL_b=13$ and their angular separation is $\zeta=21\degree$. For the Hellings and Downs curve, the fractional difference between the magnitude of the full expression and the Earth term only expression is only $\sim 3\%$.

\section{Conclusions}
\label{sec:conclusion}
In this paper we have allowed the pulsar distances in a PTA to vary in the evaluation of the magnitude of the isotropic, dipole and quadrupole overlap reduction functions. For the first time, an in-depth study of the behavior of the pulsar term has been carried out, focussing on the strong pulsar term regime-- when pulsar pairs are separated by a few radiation wavelengths or less, see Figs \ref{fig:00-11}, \ref{fig:fig22}. Although the stochastic GW background is expected to be largely isotropic at $10^{-9}$~Hz, we have included the anisotropic overlap reduction functions for completeness, and included a new study of their features in Sec \ref{sec:ETfeatures}. 

In Sec \ref{s:distance}, we found that in a $f\sim10^{-9}$~Hz stochastic GW background, and for pulsars $100$~pc from the solar system barycenter, the pulsar term is the most important for equidistant pulsars. We calculated the fractional differences between the full and Earth-term-only ORFs, reported in Table \ref{table1}, for ORFs up to $l=2$. Interestingly, we find that the most significant fractional differences between the full and Earth-term only ORFs are found in the anisotropic ORFs, with the maximum value of the magnitude of the ORF achieved at non-zero pulsar separations. For example, for $^{(ab)}\Gamma^2_2(fL, \zeta)$, the maximum fractional difference between the full and Earth-term ORF is 188 for pulsars separated by $3.1\Degs$.

More relevant to current stochastic GW background searches is the fractional difference between the magnitude of the full and Earth-term-only isotropic ORF, which is most important for pulsars separated by less than a radiation wavelength, see Table \ref{table1}, Figs \ref{fig:mag00}, \ref{fig:contour00}. Therefore a Taylor Series expansion of the pulsar term was calculated in Sec~\ref{s:pulsarterm}, and this expression can be readily input into GW data analysis pipelines. We find the approximation should be used for pulsar pairs separated by $\zeta\leq2.3\Degs$, cf. Eq. \eqref{eq:approxZeta}. In this range, the Taylor series expansion closely follows the form of the full ORF.

Looking to the future, we reported the behavior of the isotropic, dipole and quadrupole ORFs when the pulsars are within a radiation wavelength of the SSB in Sec~\ref{s:closepulsar}. We found there would be strong deviations from the usual delta-function like behavior of the pulsar term, which is currently used in searches.

It is clear from this study that the Earth-term only approximation of the overlap reduction function is still very good for the {\it current} millisecond pulsar population timed by PTAs. However, as more millisecond pulsars are added to PTAs, one should be careful to check that all the conditions for using the Earth-term only overlap reductions function still hold.

\section*{Acknowledgements}
The authors would like to thank Alberto Vecchio, Will Farr, Ilya Mandel, Cees Bassa, Rutger van Haasteren and colleagues from the European Pulsar Timing Array. This research made use of Python and its standard libraries, numpy and matplotlib.

\begin{appendix}

\section{Features of the overlap reduction function in the strong pulsar term regime}
\label{sec:features}

\begin{figure*}
\centering
 \subfigure[The $Y^{0}_{2}(\theta,\phi)$ spherical harmonic]{%
            \includegraphics[width=0.35\textwidth]{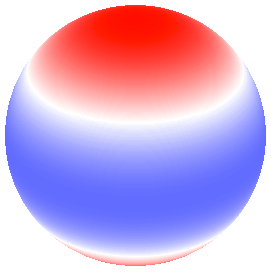}
        }%
 \subfigure[Contributions to the ORF]{%
            \includegraphics[width=0.3\textwidth]{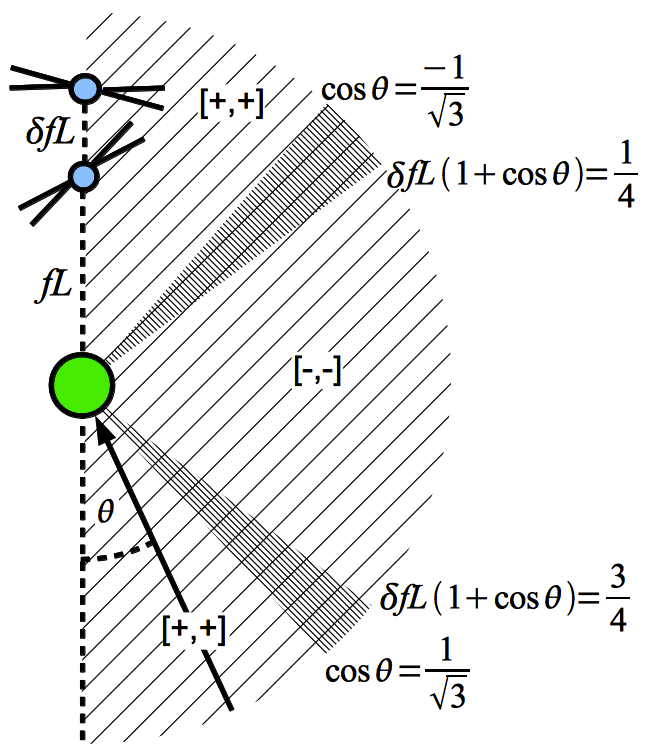}
        }%
\caption{ (a) The energy density distribution for $Y^0_2(\theta,\phi)$. The red and blue regions are positive and negative, respectively. 
(b)   The Earth (green, equivalently the SSB) is at the centre with
  the two pulsars above. The magnitude of the $^{(ab)}\Gamma^0_2(fL,\zeta)$ ORF is enhanced by small $0<\delta fL\lesssim1$ pulsar $b$ displacements, over the $\delta fL=0$ case. 
The arrow shows the direction of a GW propagating with incoming angle $\theta$. The lighter shaded regions of the
  diagram show the regions of the sky from which the
  signal will contribute positively to the ORF. The darker shaded regions will
  contribute negatively to the ORF, though their size depends
on $\delta fL$. The brackets indicate
  the [sign of the pulsar term correlation, sign of the background energy density]. }
\label{fig:quad}
\end{figure*}

Stochastic GW background searches all assume that many GWs separate pulsar pairs from each other and the SSB.
However, when the pulsars are separated by
a few radiation wavelengths or less, there is a coherent addition of the GW phase between 
neighboring pulsars, cf. Figs~\ref{fig:00-11}, \ref{fig:fig22}.
In Sec \ref{s:distance}, we probed the strong pulsar term regime by fixing pulsar $a$ and 
moving pulsar $b$ azimuthally by $\zeta fL_a$ and radially by $\delta fL=fL_b-fL_a$, see
 Fig \ref{fig:pulsar_cartoon}. Some of the contour plots in Figs~\ref{fig:00-11}, \ref{fig:fig22} showed new 
and interesting behavior in this regime, including large fractional difference between the magnitude of the ORF with respect to the Earth-term-only ORF, for pulsars separated by a few degrees, cf. Table~\ref{table1}.  

Here we explain these features by considering the interplay between the geometry of the pulsar-Earth system and its alignment with the GW energy density decomposed over the basis of spherical
harmonics. The doubling of the ORF at $\zeta=0$ is a known feature, cf. Eq.~\eqref{eq:RfullKappa}. 
The geometry in Fig \ref{fig:geometry} is used.
%In the following, pulsar $a$ is aligned with the $z$-axis.

Take for example the $Y^{0}_{2}(\theta,\phi)$ spherical harmonic. 
In Fig \ref{fig:quad}, we show that it has both positive and
negative regions which contribute positively and
negatively to the ORF respectively. The product of the
positive/negative correlation introduced by the pulsar term (which is
in turn a function of the separation of the pulsars and the direction
of the incoming wave, $\theta$) and the sign of the spherical
harmonic in a particular region of the sky, gives the overall sign of
the ORF in that region. By studying how the correlated phase changes
interact with GW energy density distribution, we will gain
some insight into the general features of the strong pulsar-term
regime.

First we examine how moving pulsar $b$ in the $z$-direction affects
the ORF in the strong pulsar term regime. When the pulsars are
separated by $\delta fL(1+\cos\theta)\leq 0.25$ the pulsar terms
introduce a positive correlated phase change. This comes from
  considering the difference in the number of GWs that the pulse from
pulsar $b$ will traverse as compared to the pulse from $a$. If this
is less than 1/4 of a radiation wavelength, the pulsar terms will be correlated.
Since the pulsars are embedded in a $Y^{0}_{2}(\theta,\phi)$-type GW background, 
sign of the GW energy density in the $\cos^{-1} (-1/\sqrt{3}) < \theta < \pi$
region is also positive. Therefore the sign of the ORF here is
positive. This region is denoted by the topmost [+,+] in Fig \ref{fig:quad}.

The pulsar terms are again positively correlated when $0.75< \delta fL
(1 + \cos\theta)<1.25$, i.e. the pulses from the 2 pulsars differ in
  the number of GWs they traverse by between 3/4 and 5/4 of a wavelength.
Moreover,  when $\theta < \cos^{-1}(1/\sqrt{3})$, the
contribution from the $Y^{0}_{2}(\theta,\phi)$ distributed GW energy density is
also positive. This region is denoted by the lower
[+,+].

When pulsar $b$ is between $0.25 \leq \delta fL\leq 0.75$ radiation wavelengths from
$a$, the pulsar term phases will be anti-correlated. However, this
region coincides with the region where the GW energy density is also
negative, and therefore the overall contribution to the ORF is positive. 
This region is denoted by [-,-]. However,
  for overlapping pulsars, or $\delta fL=0$, the pulsar terms would be positively
  correlated. In this case, the aforementioned region would contribute negatively to the ORF.
This explains why some large ORF values are observed for
pulsars which are separated by a small $\delta fL$ or equivalent angle, cf. Table \ref{table1}, though it should
be noted that for the particularly favorable setup that resembles the
region sizes shown in Fig \ref{fig:quad} would require $\delta fL \approx 1/2$.

Analogous arguments hold when moving pulsar $b$ azimuthally,
separating the pulsars by $\zeta fL_a$ radiation wavelengths, though
the difference in the number of GWs the pulses from the 2 pulsars traverse is now given
by $\approx fL \zeta \sin \theta$. These arguments are also important for explaining features seen in the other
anisotropic ORFs, though not always so straightforwardly as the energy density
distributions do not all have rotational symmetry around the $z$-axis.

\end{appendix}

\end{document}